\documentclass{emulateapj}
\usepackage{xspace}
\usepackage{amsmath}
\usepackage{hyperref}
\usepackage{framed} 
\usepackage{txfonts}
\usepackage{epstopdf} 
\newcommand{\kmsmpc}{\>{\rm km}\,{\rm s}^{-1}\,{\rm Mpc}^{-1}}

\newcommand{\kpch}{\>{h^{-1}{\rm kpc}}}
\newcommand{\mpch}{\>h^{-1}{\rm {Mpc}}}
\newcommand{\gpch}{\>h^{-1}{\rm {Gpc}}}

\newcommand{\Msunh}{\>h^{-1}\rm M_\odot}

\newcommand{\beq}{\begin{equation}}
\newcommand{\eeq}{\end{equation}}

\newcommand{\drm}{{\rm d}}

\newcommand{\vecb}[1]{{\bf #1}}
\newcommand{\pinf}{P_\infty}

\def\gtsima{$\; \buildrel > \over \sim \;$}
\def\ltsima{$\; \buildrel < \over \sim \;$}
\def\prosima{$\; \buildrel \propto \over \sim \;$}
\def\gsim{\lower.7ex\hbox{\gtsima}}
\def\lsim{\lower.7ex\hbox{\ltsima}}
\def\simgt{\lower.7ex\hbox{\gtsima}}
\def\simlt{\lower.7ex\hbox{\ltsima}}
\def\simpr{\lower.7ex\hbox{\prosima}}

\newdimen\hssize
\newdimen\hdsize
\begin{document}

\title[The overdensity of the Friends-of-Friends Halos]
       {The overdensity and masses of the Friends-of-Friends Halos and
       universality of halo mass function}
\author{Surhud More\altaffilmark{1,2},
       Andrey V. Kravtsov\altaffilmark{2,3,4},
       Neal Dalal\altaffilmark{5}, Stefan Gottl\"ober\altaffilmark{6}
       }
\altaffiltext{1}{{\tt surhud@kicp.uchicago.edu}}	   
\altaffiltext{2}{Kavli Institute for Cosmological Physics and Enrico
  Fermi Institute, The University of Chicago, Chicago, IL 60637 USA} 
\altaffiltext{3}{Department of Astronomy \& Astrophysics, The
  University of Chicago, Chicago, IL 60637 USA} 
\altaffiltext{4}{Enrico Fermi Institute, The University of Chicago,
Chicago, IL 60637}
\altaffiltext{5}{Canadian Institute for Theoretical Astrophyics,
University of Toronto, 60 St. George St., Toronto, Ontario M5S 3H8,
Canada} \altaffiltext{6}{Astrophysikalisches Institut Potsdam, An der
Sternwarte 16, 14482 Potsdam, Germany}





\begin{abstract}
The friends-of-friends algorithm (hereafter, FOF) is a percolation
algorithm which is routinely used to identify dark matter halos from
N-body simulations. We use results from percolation theory to show
that the boundary of FOF halos does not correspond to a single
density threshold but to a range of densities close to a critical
value that depends upon the linking length parameter, $b$.  We show
that for the commonly used choice of $b=0.2$, this critical density is
equal to $81.62$ times the mean matter density. Consequently, halos
identified by the FOF algorithm enclose an average overdensity which
depends on their density profile (concentration) and therefore changes
with halo mass contrary to the popular belief that the average
overdensity is $\sim$180. We derive an analytical expression for the
overdensity as a function of the linking length parameter $b$ and the
concentration of the halo. Results of tests carried out using
simulated and actual FOF halos identified in cosmological simulations
show excellent agreement with our analytical prediction.  We also find
that the mass of the halo that the FOF algorithm selects crucially
depends upon mass resolution. We find a percolation theory motivated
formula that is able to accurately correct for the dependence on
number of particles for the mock realizations of spherical and
triaxial Navarro-Frenk-White halos. However, we show that this
correction breaks down when applied to the real cosmological FOF
halos due to presence of substructures. Given that abundance of 
substructure depends on redshift and cosmology, we expect that 
the resolution effects due to substructure on the FOF mass and halo mass
function will also depend on redshift and cosmology and will be difficult
to correct for in general. Finally, we discuss the
implications of our results for the universality of the mass function.
\end{abstract}




\section{Introduction}
\label{sec:intro}

Over the last three decades, cosmological simulations have been
playing an ever increasing role in testing cosmological structure
formation models against observations using statistics that can be
reliably measured in both. Given that most of the available
observational information is about virialized peaks in the overall
matter distribution, identification of corresponding virialized peaks,
or {\it halos}, in simulations is of critical importance.

A number of automated halo finding algorithms have been developed over
the years \citep[e.g.,][and references therein]{knebe_etal11}. One of
the most popular of these is the ``Friends-Of-Friends'' (hereafter,
FOF) algorithm which uniquely defines groups that contain all
particles separated by distance less than a given linking length,
$b\bar{l}$, where $\bar{l}$ is the mean interparticle separation in
simulations and $b$ is a free parameter of the algorithm.  The FOF
algorithm is commonly applied both to identify
groups of galaxies in redshift catalogs
\citep{huchra_geller82,press_davis82,einasto_etal84,eke_etal04,berlind_etal06}
and virialized halos in cosmological simulations
\citep{einasto_etal84,davis_etal85,frenk_etal88,lacey_cole94,klypin_etal99,jenkins_etal01,warren_etal06,gottloeber_yepes07}.

 An attractive feature of the FOF algorithm is its simplicity: the
result depends solely on the linking length in units of the mean
interparticle separation, $b$. The FOF algorithm does not assume any
particular halo shape and can therefore better match the generally
triaxial mass distribution in halos forming in hierarchical structure
formation models. In addition, studies over the last decade indicate
that the appropriately parameterized mass function of FOF halos is
universal for different redshifts and cosmologies at least to $\sim
10\%$, although real systematic variations of $\lesssim 10\%$ do exist
\citep{jenkins_etal01,white02,evrard_etal02,hu_kravtsov03,warren_etal06,reed_etal07,lukic_etal07,tinker_etal08,bhattacharya_etal10,crocce_etal10,courtin_etal10}. 
Mass function of halos identified using the spherical overdensity (SO)
algorithm, on the other hand, exhibits considerably larger differences for different cosmologies and redshifts \citep{white02,tinker_etal08}. Given the
importance of the halo mass function in interpreting observed counts
of galaxies and clusters, it is interesting to understand the origin
of deviations from universality, the role of mass definition, and
differences between mass functions defined with the FOF and SO halo
finders
\citep[e.g.,][]{audit_etal98,jenkins_etal01,white01,white02,tinker_etal08,lukic_etal09}. This,
in turn, requires good understanding of properties of the
FOF-identified groups. For example, a recent study by
\citet{courtin_etal10} shows that the degree of universality depends
sensitively on the choice of the linking length parameter $b$.

 One could expect that for a given value of $b$, the FOF algorithm
defines the boundary of a halo as corresponding to a certain isodensity
surface, at least in the limit of large number of particles.
\citet{frenk_etal88} indicate that the overdensity (defined with respect to the mean density of the universe: $\delta=\rho/\bar{\rho}-1$) of this surface is $\delta_{\rm fof}\approx 2b^{-3}$. \citet[][see also \citeauthor{summers_etal95} \citeyear{summers_etal95} and \citeauthor{audit_etal98} \citeyear{audit_etal98}]{lacey_cole94} quote a value four times smaller, of 
$\delta_{\rm fof}=3/(2\pi b^3)\approx 0.48b^{-3}$, 
corresponding to the local overdensity of two particles within a sphere
of radius $b$. Clearly, such local overdensity is the absolute minimum
overdensity that should be sampled by the particles of an FOF
halo. For the most commonly used value of $b=0.2$ this corresponds to
a local overdensity of $\delta_{\rm fof}\approx 60$, which for an
isothermal density profile, $\rho(r)\propto r^{-2}$, corresponds to an
enclosed overdensity of $3\delta_{\rm fof}\approx 180$. This value is
close to the virial overdensity predicted by the spherical collapse
model in the Einstein-De Sitter cosmology and is usually regarded as a
justification for using $b=0.2$ in analyses of simulations.

More recently, \citet{warren_etal06} have noted that their experiments
on Poisson realizations of isothermal halos indicate that the
FOF algorithm identifies the boundary at an overdensity $\delta_{\rm
fof}\approx 74$, which corresponds to an enclosed overdensity of
$\approx 280$ rather than the canonical value of 180. Indeed, they
report that direct measurements of internal overdensities of the FOF
halos in their cosmological simulations identified with $b=0.2$ range
from $\sim 200$ for largest simulation boxes to $\sim 400$ for the
smallest boxes. Given that small boxes resolve predominantly smaller
mass halos compared to larger boxes, this result hints that the
internal overdensity of the FOF halos is actually mass dependent. 

Given that the FOF algorithm identifies boundary at a {\it local}
overdensity and halos are described by an \citet[][hereafter
NFW]{nfw97} profile with mass-dependent concentration, this result is
not surprising. However, concentration also strongly depends on
cosmology and
redshift \citep[e.g.,][]{bullock_etal01,zhao_etal03b,zhao_etal09},
which immediately implies that the internal overdensity of FOF halos
identified with a given value of $b$ is also redshift and cosmology
dependent. Interpretation of the FOF halo mass function and other
statistics is therefore not trivial. For example, Halo Occupation
Distribution (HOD) models typically assume that halos are defined
within a spherical radius enclosing a well-defined overdensity. Also,
creating mock galaxy catalogs by assigning galaxies to FOF halos
requires knowledge of the internal halo overdensities in order to
model the target galaxy bias properly.

In this study we present a detailed analysis of the halo boundary and
the corresponding overdensity selected by the FOF algorithm with a given
linking length $b$ based on random particle realizations of
spherical NFW halos. We also present an analytical interpretation of the
results of these experiments and compare its predictions to
overdensities of FOF halos in cosmological simulations. We show that
the boundary of the FOF halos corresponds not to a single local
overdensity, but to a range of overdensities around a characteristic
value that can be understood on the basis of percolation
theory. For the commonly used value of $b=0.2$, the characteristic
local overdensity is $\delta\approx 81$, a value higher
than that quoted in previous studies. Correspondingly, the enclosed
overdensity of the FOF halos is considerably higher than thought
before and for $b=0.2$ ranges from $\sim 250$ to $\sim 600$ for
typical halo concentrations (overdensities for other values of $b$
scale as $\propto b^{-3}$).

The paper is organized as follows. In \S~\ref{sec:od} we present tests
of the FOF algorithm on Monte Carlo realization of idealized spherical
NFW halos and show explicitly that 1) the boundary of FOF halos does
not correspond to a single local overdensity, but rather to a range of
overdensities, 2) the enclosed overdensities of the FOF halos are
significantly larger than commonly thought and depend on
concentrations of halos and thus on mass, redshift, and cosmology. In
\S~\ref{sec:fofod} we develop a simple analytic model that
encapsulates results of the Monte Carlo experiments of \S~\ref{sec:od}
(see also the Appendix for interpretation of these results in the
context of percolation theory) and present tests of this model against
results of cosmological simulations. In \S~\ref{sec:mf} we discuss
implications of our results for the universality of halo mass
function. In \S~\ref{sec:masses}, we interpret results for idealized
realizations of NFW halos in the context of percolation theory and
present an accurate formula describing the dependence of the FOF mass on mass resolution based on this theory. In \S~\ref{sec:masses} we also consider real halos extracted from cosmological simulations of a $\Lambda$CDM cosmology and show that substructure present in real halos makes behavior of the FOF masses with resolution even more complicated. Finally, we summarize our results and conclusions in
\S~\ref{sec:conc}. In the Appendix, we review the basics of the percolation theory and demonstrate how the boundary of the FOF halos and their mass can be understood and predicted in its context.

\section{Tests with Monte Carlo realizations of spherical NFW halos}
\label{sec:od}

To explore the boundary of the FOF halos and their enclosed
overdensities, we follow the approach of \citet{lukic_etal09} and consider
Monte Carlo realizations of idealized spherical halos. We assume that
the
internal density distribution of the halos is described by the NFW
density profile \citep{nfw97}:
\begin{equation}
n(r) = \frac{A}{\left( r/r_{\rm s}\right) \left( 1 + r/r_{\rm s}
\right)^2} \,,
\label{eq:nfw}
\end{equation}
which is a reasonable approximation to density profiles of halos
formed in CDM cosmologies. Here, $r_s$ denotes the scale radius. The
boundary of a halo is usually defined with respect to the radius
$R_{\Delta}$ that encloses internal overdensity $\Delta$ with respect
to the mean density of the universe.  The radii $r_s$ and $R_{\Delta}$
are related via the concentration parameter
$c_{\Delta}=R_{\Delta}/r_s$.  The normalization, $A$, is then given by
\begin{equation}
A = \left(\frac{N_{\Delta}}{4\pi R^3_{\Delta}}\right)
\frac{c_{\Delta}^3}{\mu(c_\Delta)}\,,
\label{eq:nfwnorm}
\end{equation}
where $N_{\Delta}$ is the number of particles within $R_{\Delta}$ and
the function $\mu(x)$ is given by
\begin{equation}
\mu(x)=\ln(1+x)-\frac{x}{(1+x)}\,.
\end{equation}

For the Monte Carlo realizations presented in this section, we assume
concentration of $c_{\Delta}=10$. We generalize our results for other concentrations in the
following section. We generate such realizations with varying number
of particles, $N_p$, and mean interparticle separation, $\bar{l}$.
The latter can be expressed in terms of the radius $R_{\Delta}$ and
the number of particles $N_{\Delta}$ as
\begin{equation}
\bar{l}=\left[ \frac{4\pi\,R_{\Delta}^3}{3}\frac{\Delta}{N_{\Delta}}
\right]^{1/3} \,.
\end{equation}
As the
boundary that the FOF algorithm will select is not known a priori, we
conservatively generate particle distribution up to the radius of
$2R_{\Delta}$.

Without loss of generality, we use $\Delta=180$, one of the most
commonly used mass-defining overdensities, and generate a series of
halo realizations with $N_{180}$ varying from $10^7$ to $100$
particles. To reduce Poisson noise, for small $N_{180}$ we generate
multiple realizations and average over them. We use $10$, $100$,
and $1000$ realizations for halos with $10^4$, $10^3$, and $100$
particles, respectively.  As the particle distribution extends
up to $2R_{180}$, the actual number of particles used in each of the
realizations is larger than $N_{180}$ roughly by a factor of $1.4$.
We run the FOF halo finder on each of the halo realizations with a linking
length equal to $0.2\,\bar{l}$. The algorithm links particles with
each other if they are closer than the linking length. In what
follows, we restrict our attention to the largest group that is found
by the FOF algorithm. 

\begin{figure}[tc]
\centering
\includegraphics[scale=0.9]{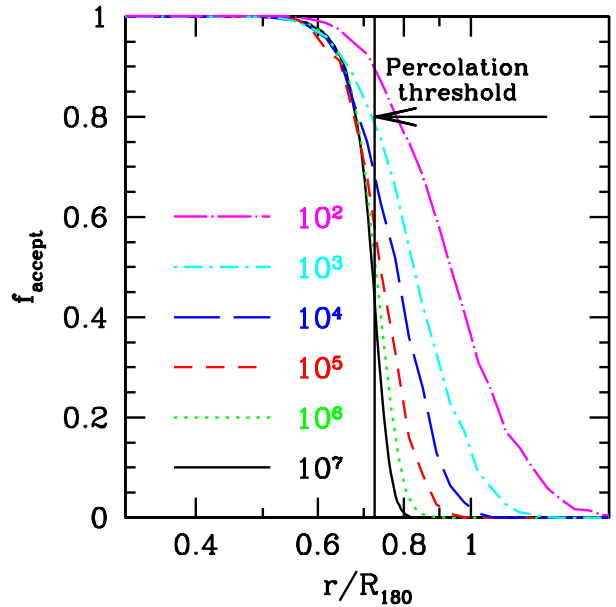}
\caption{ 
The fraction of particles that are joined into the largest group by
the FOF algorithm with $b=0.2$ as a function of the radius (in units
of the radius $R_{180}$) enclosing the mean overdensity $\Delta=180$ for Monte Carlo realizations of spherical NFW halos with
varying number of particles, $N_{180}$ (lines of different style and color, as
indicated in the legend). The vertical solid line marks the radius at
which the density equals the critical threshold for percolation (eqs.~\ref{eq:ncrit} and \ref{eq:dfof}).  }
\label{fig1}
\end{figure}

\begin{figure}[tc]
\centering
\includegraphics[scale=0.9]{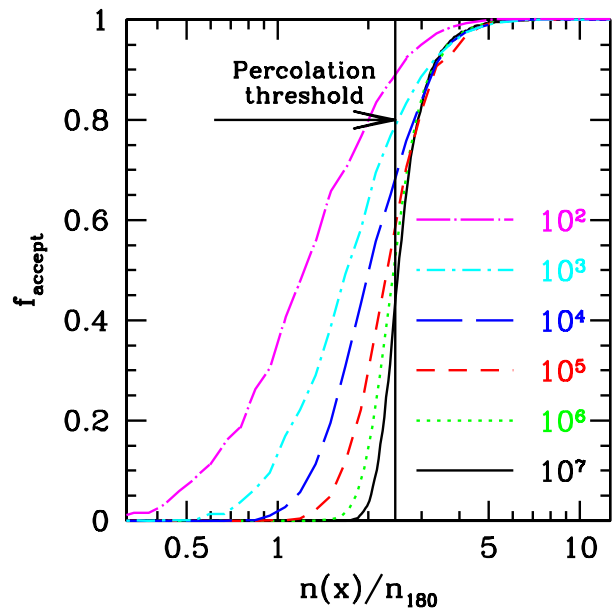}
\caption{ 
      Same as Figure~\ref{fig1} but as a function of the local number
      density (calculated analytically using the position of the
      particle), in units of the local number density at $R_{180}$.
  }
\label{fig2}
\end{figure}

\begin{figure}[t]
\centering
\includegraphics[scale=0.9]{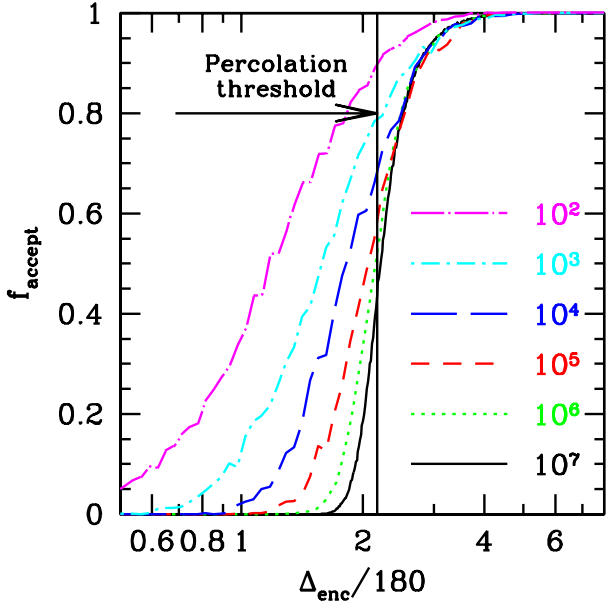}
\caption{ 
      Same as Figure~\ref{fig1} but as a function of the average
      enclosed overdensity, $\Delta_{\rm enc}$, normalized to overdensity of 180.
  }
\label{fig3}
\end{figure}

Figures~\ref{fig1},\ref{fig2}, and ~\ref{fig3} show the fraction of
particles in a Monte Carlo halo, $f_{\rm accept}$, that are grouped
into the central halo by the FOF algorithm at a given radius as a
function of radius, local density, and enclosed overdensity,
respectively. Although we generate realizations of spherically
symmetric halos with no physical substructure, the figures show that
the boundary of the FOF-identified halos is not sharp. The particles
joined into the FOF group span a range of radii and overdensities. The
``fuzziness'' of the boundary increases dramatically for realizations
with the smallest number of particles. Note, however, that even for
realizations with millions of particles, $f_{\rm accept}$ as a
function of radius or overdensity does not converge to a step function,
but rather converges to a well-defined shape spanning a range of
radii. This implies that the
boundary selected by the FOF algorithm is inherently fuzzy.

Figure~\ref{fig2} also clearly shows that the local overdensity of
majority of particles within the fuzzy FOF boundary is larger than
$n_{180}$. Correspondingly, the mean enclosed overdensity within this
boundary is also much larger than $180$, contrary to what is usually
assumed for $b=0.2$ linking length (Fig.~\ref{fig3}).

The particles that are joined into an FOF group depend upon the percolation
properties of the particle distribution. In the Appendix, we show that the behavior
of $f_{\rm accept}$ as a function of radius and overdensity
demonstrated by Figures~\ref{fig1}-\ref{fig3} can be understood in the
framework of percolation theory. For example, percolation theory
predicts that for a uniform particle distribution percolation (i.e.,
formation of a group spanning the entire region) should occur at the
local number density equal to a critical value of 
\begin{equation}
n_{\rm crit}=\frac{n_c}{(b \bar{l})^{3}}\,,
\label{eq:ncrit}
\end{equation}
This corresponds to the local overdensity (with respect to the mean
density $\bar{n}=\bar{l}^{-3}$) of 
\begin{equation}
\delta_{\rm crit}\equiv \frac{n_{\rm crit}}{\bar{l}^{-3}}-1=n_c\,b^{-3}-1.
\label{eq:dfof}
\end{equation}
 Here $n_c$ is a
universal constant that arises in the percolation problem of spheres
that follow a Poisson distribution. The value of this constant has
been calibrated via extensive Monte Carlo experiments \citep{lorenz_ziff01}: 
\begin{equation}
n_c=0.652960 \pm 0.000005\,.
\label{eq:ncritval}
\end{equation}

We can expect that the boundary of FOF halos should approximately
correspond to $n_{\rm crit}$ because percolation across a radial bin
will be inhibited at smaller densities.  For our choice of $b=0.2$,
this corresponds to $n_{\rm crit}=81.62\,\bar{l}^{-3}$, i.e. local
overdensity $\delta_{\rm crit}=80.62$. This overdensity is shown by the
vertical line in Figure~\ref{fig2}, while vertical lines in
Figures~\ref{fig1} and \ref{fig3} show the corresponding radius and
enclosed mean overdensity. The figures show that percolation threshold
does indeed predict a characteristic overdensity and radius roughly in
the middle of the FOF boundary range. In the Appendix, we show that
percolation theory also explains the shape of $f_{\rm accept}$ as a
function of radius and overdensity for $n>n_{\rm crit}$, and the
increase in the fuzziness of the boundary with decreasing number of
particles used.

For our immediate purposes, however, we can consider the empirical
results of our Monte Carlo tests for the overdensities of the FOF
halos. In the next section, we present a simple analytic model that
describes this overdensity as a function of linking length $b$ and
halo concentration $c$.

\begin{figure}[t]
\centering
\includegraphics[scale=0.9]{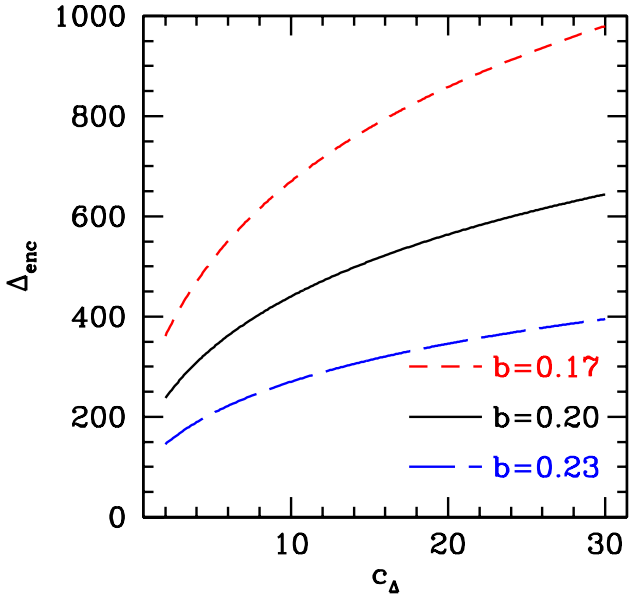}
\caption{The overdensity of halos as a function of the concentration
of halos selected by the FOF algorithm for three representative
values of the linking length $b=0.17$, $0.20$, $0.23$ shown by the
short-dashed, solid, and long-dashed lines, respectively. 
  }
\label{fig4}
\end{figure}
%

\section{Concentration dependence of the enclosed FOF Overdensity}
\label{sec:fofod}

\subsection{Analytical model}

In the previous section we showed that the boundary of the FOF
algorithm corresponds to a wide range of local overdensities (with the
width of the range dependent on the number of particles in a halo)
around a characteristic local density $n_{\rm crit}=n_{\rm
c}\,(b\,\bar{l})^{-3}$ or the corresponding local overdensity
$\delta_{\rm crit}\equiv n_{\rm crit}/\bar{n}-1=n_{\rm c}\,
b^{-3}-1$. For the commonly used value of the linking length parameter
$b=0.2$, $\delta_{\rm crit}=80.62$. Given the characteristic local
overdensity at the boundary, it is straightforward to derive an
analytical expression for the average enclosed overdensity assuming
that halos have NFW density profiles.

Let us denote the number of particles selected by the FOF algorithm as
$N_{\Delta}$, and the effective spherical radius enclosing these
particles as $R_{\Delta}$, where $\Delta$ is the overdensity of the
FOF halo which we wish to determine. Evaluating the number density at
$R_{\Delta}$ using equations~\ref{eq:nfw} and ~\ref{eq:nfwnorm}, and
equating it to the critical number density, $n_{\rm crit}$ yields
\begin{equation}
\left( \frac{ N_{\Delta}}{ 4\pi R_{\Delta}^3 } \right)
\frac{c_{\Delta}^3}{\mu(c_\Delta)}  \frac{1}{c_{\Delta}(1+c_{\Delta})^2}  =
n_{\rm c} (b\,\bar{l})^{-3}\,.
\end{equation}
Note that here $c_{\Delta}\equiv R_{\Delta}/r_s$ is the concentration
defined with respect to $R_{\Delta}$. 

The enclosed overdensity, $\Delta$, of the halo is then given by
\begin{eqnarray}
\Delta &=& \left( \frac{3N_{\Delta}}{ 4\pi R_{\Delta}^3 \bar{l}_{\rm}^{-3} } \right) - 1 \\
  &=& 3\,n_{\rm c} b^{-3} \frac{ \mu(c_\Delta)
  (1+c_{\Delta})^2 }{c_{\Delta}^2} - 1\,.
\label{eq:od}
\end{eqnarray}
This explicitly shows that the overdensity of an FOF halo depends not
only upon the linking length parameter, $b$, but also upon its
concentration. In Figure~\ref{fig4}, we show the average
FOF halo overdensity as a function of the concentration, $c_{\Delta}$,
for three representative values of $b$.

Note that one needs to know the concentration-mass relation to predict
the overdensity of halos as a function of the FOF halo mass. The
concentration of halos depends upon the radius of the halo (and hence
the overdensity definition). The concentration and the average
overdensity of FOF halos as a function of their mass can be
calculated using the following steps. (i) As a first guess, we assume
that FOF halos have a certain overdensity (say $\Delta_i$) with
respect to the background. (ii) We use the concentration-virial mass
relation given by \citet{zhao_etal09}\footnote{ \citet{zhao_etal09}
calibrate concentration-mass relation for concentration and masses
defined with respect to the radius enclosing virial overdensity,
$\Delta_{\rm vir}$. } and convert it to a   concentration-mass 
relation for halos with overdensity $\Delta_i$. (iii) This concentration is used to find a
new overdensity using Eq.~\ref{eq:od}.  We repeat steps (ii) and (iii)
until we converge to a value of overdensity (and concentration). 

Note that since the concentration of a halo depends on cosmology,
redshift, and halo mass, the enclosed overdensity of halos selected by
the FOF algorithm also depends upon cosmology, redshift, and
mass. Furthermore, even for a fixed cosmology, redshift, and mass, halo
concentrations exhibit substantial scatter and we can therefore expect
a corresponding scatter in enclosed overdensities. We will consider
these dependencies and scatter in the next section, where we compare
the predictions of equation~\ref{eq:od} to overdensities of FOF halos
identified in cosmological simulations.

\subsection{Comparison with cosmological simulations}

To test the simple model presented in the previous section, we compare
predictions of equation~\ref{eq:od} with actual overdensities of halos
identified in dissipationless cosmological simulations of the
$\Lambda$CDM model. Halos have been identified using the FOF algorithm
with different linking lengths $b$ and at different redshifts in two
cosmological simulations of the same flat $\Lambda$CDM cosmology: the
matter and baryon density in units of the critical density
$\Omega_{\rm m}=1-\Omega_{\Lambda}=0.27$ and $\Omega_{\rm b}=0.0469$,
the Hubble constant $h=H_0/(100 \kmsmpc)=0.70$, the rms amplitude of
linear fluctuations in spheres of radius $8h^{-1}$~Mpc
$\sigma_8=0.82$, and the power law slope of the initial power
spectrum, $n_s=0.95$. 

\begin{figure}[t]
\includegraphics[scale=0.9]{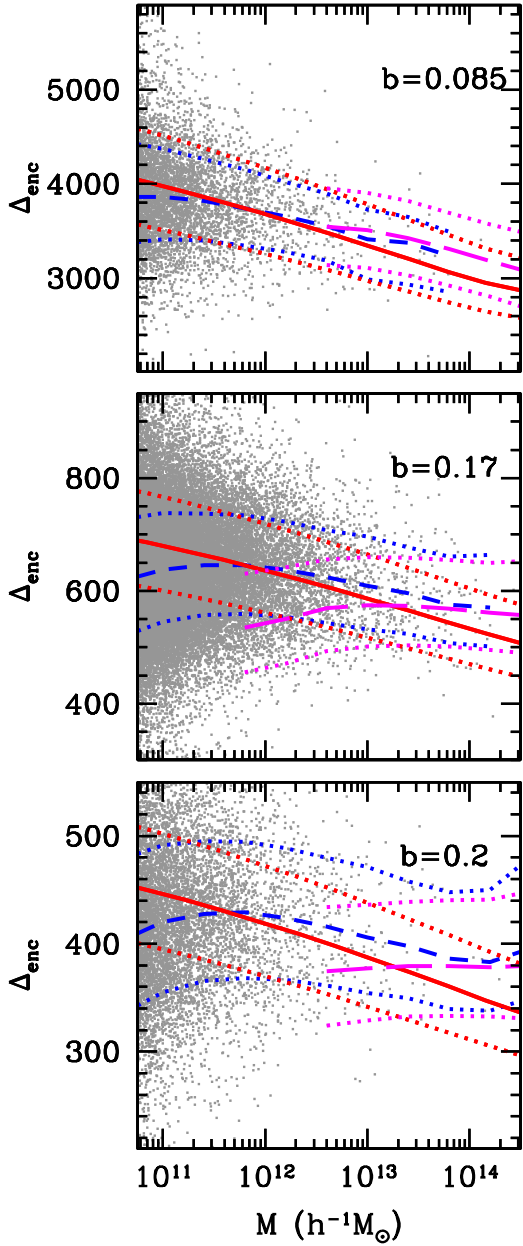}
\caption{Enclosed overdensities of the FOF halos identified with
linking lengths $b=0.085$, $0.17$, and $0.20$ in the Bolshoi and
MultiDark simulations. In each panel, the dashed lines show the median
overdensity, while the dotted lines show the 16 and 84 percentiles of
the distribution. The blue and purple lines correspond to the results
of the Bolshoi and MultiDark simulations, respectively. The grey
points show halos from the Bolshoi simulation (the MultiDark halos are
not shown for clarity). The red solid lines show the prediction of our
model given by eq.~\ref{eq:od} and concentration--mass relation of
\protect\citet{zhao_etal09}.  The red dotted lines show the rms
scatter predicted by the model, if we assume a scatter of $0.14$ dex of
concentrations at a given mass.  }
\label{fig5}
\end{figure}
The first is the Bolshoi simulation of a cubic volume of size $L_{\rm
B}=250\mpch$, described in detail in \citet{Klypin2010}, while the
second is the MultiDark simulation of volume size $L_{\rm MD}=1\gpch$
(Prada et al., in preparation)\footnote{Data from both simulations are publicly available at http://www.multidark.org/MultiDark/ .}. Both simulations followed the evolution of
$2048^3$ particles, which corresponds to particle masses of
$1.36 \times 10^8 \Msunh$ and $8.72\times 10^9 \Msunh$ for the Bolshoi
and MultiDark simulations, respectively. The peak spatial resolution
was $1\kpch$ and $7\kpch$ in these simulations, respectively.

The FOF algorithm used to identify halos in these simulations is based
on the minimal spanning tree and is described in Knebe et
al. (2011). Given that the shape of the FOF halos can be arbitrary and
rather complicated, measurement of their volume is not trivial. We
estimate the volume employing the following procedure. For each FOF
halo, the three dimensional distribution of particles is projected
onto a two dimensional plane perpendicular to one of the coordinate axis
(e.g., the $x$-axis in the following). A grid of cells of size
$s=b\,\bar{l}$ is then overlaid on this plane. The volume occupied by
particles in each individual cell $i$ is estimated as
\begin{equation}
V_{i,x}=s^2\times (x_{\rm max}-x_{\rm min}),
\end{equation}
where $x_{\rm min}$ and $x_{\rm max}$ are the minimum and maximum $x$
coordinates of particles in the cell and $x_{\rm max}-x_{\rm min}$ is
the extent of the particle distribution along $x$. The total volume of
the halo, $V_{x}$ is calculated as a sum over all cells containing
particles $V_x=\Sigma_i V_{i,x}$.  This procedure is repeated for the
other two axes and the final halo volume is assumed to be the maximum
of $V_x$, $V_y$, and $V_z$.

The procedure used for estimating the volume roughly approximates the {\it convex hull} algorithm.\footnote{{\tt http://en.wikipedia.org/wiki/Convex$\_$hull}} It is designed to avoid the pitfall of estimating volume using 3D grid as a sum of cells containing particles. Such estimate leaves many empty cells within the halo unaccounted for. Moreover, such method does not converge to a well-defined volume value as the 3D grid cell size is varied. 

Figure~\ref{fig5} shows overdensities of individual FOF halos
selected from simulations as a function of the FOF halo mass selected
using different linking length parameters. The three panels show
results for FOF with linking lengths $b=0.085$, $b=0.17$ and
$b=0.2$. In each panel, the dashed lines show the median overdensity
as a function of halo mass, while the dotted lines show the 16 and 84
percentiles of the distribution. The blue (short-dashed) and purple
(long-dashed) lines correspond to the results of the Bolshoi and
MultiDark simulations, respectively.  The red solid lines show the
prediction for the overdensity of FOF halos as a function of halo
mass given by eq.~\ref{eq:od} and concentration--mass relation
of \citet{zhao_etal09}.  The red dotted lines show the rms scatter
predicted by the model, if we assume scatter of $0.14$ dex of
concentrations at a given mass, as measured in cosmological
simulations \citep[e.g.,][]{wechsler_etal02}.

The figure shows that the simple model of equation~\ref{eq:od}
captures the median overdensities of FOF halos at these different
linking lengths rather well. The scatter of overdensities in simulated
halos is also consistent with the scatter expected for the scatter in
concentrations. The mass dependence of $\Delta$ is 
qualitatively consistent in the model and simulations, except perhaps at the
smallest and largest masses. At small masses overdensities of
simulated halos exhibit a downturn in both the Bolshoi and MultiDark
simulations. The masses at which the downturn occurs are different in
the two simulations. This downturn is due to
the percolation properties of halos represented by small particle
numbers, as we discuss in more detail in \S~\ref{sec:masses} below and in the
Appendix.  Note, for
example, that the downturn shifts to smaller masses for smaller values of
$b$ (i.e., larger local particle densities at the boundary) and almost
entirely outside the shown mass region for $b=0.085$. The
overdensities of simulated halos also exhibit a somewhat weaker trend
with mass than predicted by our model for masses $\gtrsim 5\times
10^{13}\mpch$. It is not clear what is the source of this discrepancy,
but we note that it is quite small and amounts to less than 10\%.

\begin{figure*}[t]
\includegraphics[scale=0.9]{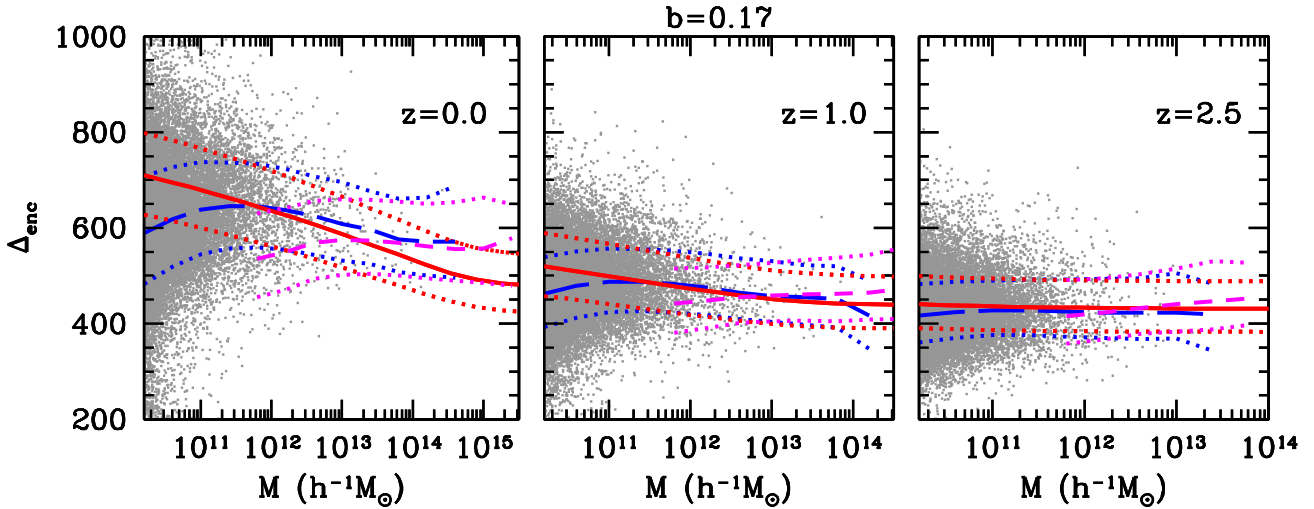}
\caption{Overdensities of the FOF halos identified with $b=0.17$ in
the Bolshoi and MultiDark simulations at redshifts $z=0.0$, $1.0$, and
$2.5$ along with median and scatter (red solid and dotted line)
predicted by our model (eq.~\ref{eq:od}). The line types and colors
are as in Figure~\ref{fig5}.
  }
\label{fig6}
\end{figure*}

Figure~\ref{fig6} shows overdensities of the FOF halos identified with
$b=0.17$ at redshifts $z=0.0$, $1.0$, and $2.5$. The evolution of
overdensity predicted by the model due to the redshift evolution of
concentrations, predicted using the model
of \citet{zhao_etal09}, matches the redshift trend observed in the
simulations remarkably well. The scatter of overdensities is also well
reproduced by the scatter of concentrations at all redshifts. Note
that enclosed overdensity for this $b$ in the mass range probed by the
simulations reaches the floor value of $\approx 400-450$ by $z=2.5$,
as virial concentration of halos reaches a floor of $c_{\rm
vir}\approx 4$ \citep{zhao_etal03a,zhao_etal09}.

\section{Implications for universality of halo mass functions}
\label{sec:mf}

Our results on the enclosed overdensity of the FOF-identified halos
have important and interesting implications for the interpretation of
recent results on the universality of the halo mass function. A number
of studies have found that the halo mass function can be expressed in
a cosmology and redshift independent way as a universal function of
the peak height, $\delta_c/\sigma(M)$, where $\delta_c(z)$ is the
linearly evolved overdensity of a peak at the time of collapse in the
spherical collapse model and $\sigma(M)$ is the {\sl rms\/}
fluctuation of perturbations of mass $M$
\citep{sheth_etal01,jenkins_etal01,warren_etal06,tinker_etal08}.  

Although deviations from universal behavior have been found for both
the FOF and SO identified halos, these deviations are markedly 
smaller for the FOF mass functions
\citep[e.g.,][]{lukic_etal07,tinker_etal08,courtin_etal10}.
\citet{courtin_etal10} showed that deviations from universality are
not random but are correlated with the nonlinear virialization
overdensity, $\Delta_{\rm vir}$, expected from the spherical collapse
model for a given cosmology and redshift. In particular, they showed
that the linking length, $b_{\rm univ}$, required to minimize
deviations of the FOF mass function from universal form for a given
cosmology and redshift is correlated with the corresponding
$\Delta_{\rm vir}$ as:
\begin{equation}
\left(\frac{b_{\rm univ}}{0.2}\right)^{-3}=0.24\left(\frac{\Delta_{\rm vir}}{178}\right)+0.68.
\label{eq:bunid}
\end{equation}

This is an interesting and important result, as it indicates that
deviations from universality can be minimized if one takes into
account cosmology-dependence of virialization parameters
properly. However, as noted by \citet{courtin_etal10}, the form of
equation~\ref{eq:bunid} is different from $(b/0.2)^{-3}=\Delta_{\rm
vir}/178$, which one would expect if the FOF algorithm with $b=0.2$
would identify halos with a constant internal overdensity of $\approx
178$. This form thus begs for a physical explanation. Our results
presented in the previous sections can help explain this empirical
correlation, at least partially. First, we showed that the typical
overdensity of FOF halos identified with $b=0.2$ at $z=0$ is
significantly larger than $178$.  Second, we showed that overdensity
of FOF halos depends not only on $b$ but also on halo concentrations
(eq.~\ref{eq:od}), and thus on mass, cosmology and redshift. In light
of these results we expect that the linking length required to identify
halos enclosing a certain overdensity $\Delta$ is given by (see eq.~\ref{eq:od})
\begin{equation}
\left(\frac{b}{0.2}\right)^{-3}=\frac{\Delta+1}{244.86}\psi(c_{\Delta}), 
\label{eq:bd}
\end{equation}
where the function $\psi(c_{\Delta})$ is given by
\begin{equation}
\psi(c)=\frac{c^2}{\mu(c)\,(1+c)^2}.
\label{eq:psi}
\end{equation}
Equation~\ref{eq:bd} can thus be used to predict what linking length
is needed to identify a halo boundary enclosing virial overdensity
$\Delta_{\rm vir}$. 

\begin{figure*}[t]
\centering
\includegraphics[scale=0.6]{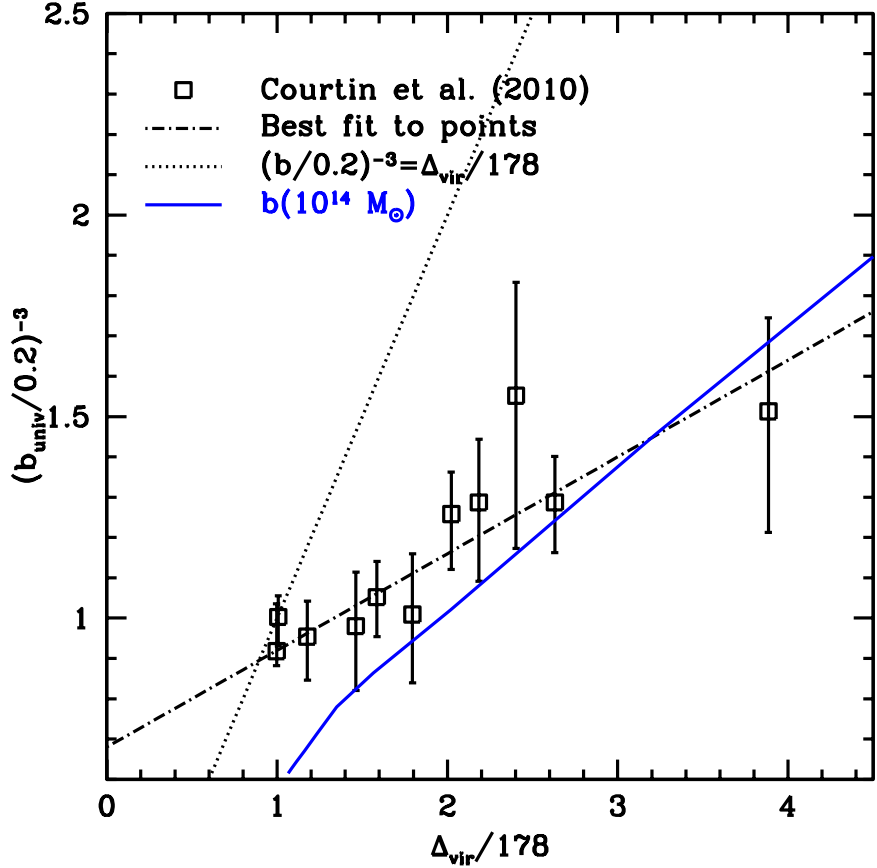}
\caption{ 
    Universality of FOF halo mass function. The linking length
    parameter, $b_{\rm univ}$ that minimizes deviations of mass
functions in different cosmologies from universal form.
    Square points and dot dashed line shows the empirical relation
    derived by \citet{courtin_etal10}. The dotted line shows the
    commonly assumed scaling between overdensity and linking length
    parameter, $b$. The solid (blue) line shows our analytical
    prediction assuming the concentration of a $10^{14} \Msunh$ halo (computed using eq.~\ref{eq:bd}, see text for details.).
  }
\label{fig:courtin}
\end{figure*}

Figure~\ref{fig:courtin} shows simulation results
of \citet{courtin_etal10} for values of $b_{\rm univ}$ as a function
of $\Delta_{\rm vir}$ (squares with error bars) and the best fit to
their results (dot-dashed line). It also shows the $b_{\rm
univ}-\Delta_{\rm vir}$ dependence given by equation~\ref{eq:bd}
(solid blue line). This line is computed assuming $c_{\rm vir}-M$
relation for a flat $\Lambda$CDM cosmology consistent with WMAP5
results given by the model of \citet{zhao_etal09} for the redshift range
from $z>2$ (where $\Omega_{\rm m}\approx 1.0$ and $\Delta_{\rm
vir}\approx 178$) to negative redshifts into the future to sample
low-$\Omega_{\rm m}$, high-$\Delta_{\rm vir}$ regime. For all
redshifts the model is computed for a fixed halo mass $M_{\rm
vir}=10^{14}\ \rm M_{\odot}$, a value representative of the mass range
probed by Courtin et al.'s simulations. As can be seen from the
figure, prediction of equation~\ref{eq:bd} is much closer to the
results of \citet{courtin_etal10} than the commonly assumed
$(b/0.2)^{-3}=\Delta_{\rm vir}/178$. Note that the slope is also
different due to dependence on concentrations via the function
$\psi(c)$. 

This implies that results of \citet{courtin_etal10} indeed indicate
that deviations from universality are largely due to the use of halo
parameters not adjusted for different virialization overdensities in
different cosmologies and redshifts. Note, however, that agreement
between our model and their results is not perfect. This could be due
to several factors. First, the virialization overdensities of halos
may be somewhat different from those expected in the spherical
collapse model, given that most halos form out of triaxial
perturbations via a complicated sequence of mergers and smooth
accretion. Second, the well-known bridging effect of the FOF halo
finder may play a role at smaller values of $\Delta_{\rm vir}$ (i.e.,
larger values of $b$). For the commonly used value of $b=0.2$ the FOF
algorithm joins $\approx 10-15\%$ of neighboring halos by
bridging at $z=0$ \citep[e.g.,][]{davis_etal85,bertschinger_gelb91,cole_lacey96,lukic_etal09},
although this fraction is likely to be higher at larger
redshifts \citep[e.g.,][]{cohn_white08}. The figure~\ref{fig:courtin},
on the other hand, shows that our model predicts that the linking
length should increase to $b\approx 0.24$ to reach $\Delta_{\rm
vir}$. We can expect that bridging will become severe for such large
linking length and would definitely affect FOF halo mass
function.\footnote{A dramatic effect of bridging on $z=10$ halo mass
function can be observed in Figure~3 of \citet{cohn_white08}, which
shows abundance of FOF halos as a function of FOF mass with $b=0.2$
and mass counted around centers of the same halos in spheres enclosing
overdensity $\Delta=180$. Although the FOF halos for $b=0.2$ should
have mean overdensities considerably larger than 180, and hence FOF
mass smaller than SO(180) mass, that figure shows that the average FOF
mass of halos of a given abundance is actually about two times larger
than their SO mass with $\Delta=180$. } The weak dependence of $b_{\rm
univ}$ on $\Delta_{\rm vir}$ for virial overdensities of $\approx
180\div 300$ may therefore reflect the fact that universality of the
FOF mass function is compromised by bridging, which prevents $b_{\rm
univ}$ from reaching lower values.

Some of the discrepancy between equation~\ref{eq:bd} and Courtin et
al. simulation results could also be due to the fact that their points
comprised simulations of different cosmologies all using the same
power spectrum and normalization $\sigma_8$ at $z=0$, while our
prediction is made for a single cosmology as a function of redshift.
Given that concentrations of halos in a given cosmology depend not
only on $\Omega_m$, but also on $\sigma_8$, results
of \citet{courtin_etal10} for $b_{\rm uni}-\Delta_{\rm vir}$ scaling
are likely not universal. For example, for cosmology with the same
$\Omega_{\rm m}$ and $\Omega_{\Lambda}$ but different values of
$\sigma_8$, halo concentrations, and hence value of $b_{\rm univ}$,
will be different but $\Delta_{\rm vir}$ will be the same.

Incidentally, the dependence of enclosed overdensity of FOF halos on
concentration could also explain why deviations of the halo mass
function from universality at different redshifts have been found to
be considerably smaller for the FOF halos identified with constant $b$
than for the SO mass function with masses defined using constant
overdensity \citep{white02,lukic_etal07,tinker_etal08,courtin_etal10}.  This
more universal behavior could, in principle, be an indication that the
FOF somehow identifies halos better related to the initial density
field or assigns mass to halos more correctly than the SO algorithm.
This would, of course, be interesting for understanding the physical
origin of the universality of the mass function.

However, given the significant bridging effect for $b\approx 0.2$ discussed
above, one should already be skeptical that some deep physics underlies
a more universal behavior of the b=0.2 FOF mass functions. In
addition, our results imply that smaller deviations of the FOF halo
mass function from universality are also due to a partial
cancellation of some of the redshift evolution of the halo mass
function by redshift evolution of halo concentrations. Indeed, for
$\Lambda$CDM models for which these deviations with redshift have been
studied, the enclosed overdensities for high-mass FOF halos at $z=0$,
when halo concentrations are relatively high, are $\sim
300-400$. These overdensities are close to the virial overdensity of
halos in the $\Lambda$CDM cosmology. At higher redshifts, however,
halo concentrations decrease as $c(M,z)\propto (1+z)^{-1}$
\citep{bullock_etal01} until they reach a floor value of $\approx 4$ \citep{zhao_etal03b, zhao_etal09}.
For $c\sim 4$, the overdensity of FOF halos should
approach $\sim 250$ (see Fig.~\ref{fig4}), which is close to the
virial overdensity at high redshifts where $\Omega_{\rm m}(z)$ is closer to
unity. The FOF overdensity thus roughly tracks the virial overdensity
in the concordance $\Lambda$CDM cosmology.  However, we stress that
this rough tracking is coincidental. This is because halo
concentrations depend on the halo formation times
\citep[e.g.,][]{wechsler_etal02,neto_etal07,zhao_etal09}, which in
turn depend on power spectrum normalization among other things. Thus,
concentrations would still evolve with redshift in the Einstein-de
Sitter $\Omega_{\rm m}=1$ cosmology, even though virial overdensity would
not. The deviations of the FOF mass function from universality would
therefore also be affected by power spectrum normalization, or any
other parameter that affects concentrations.

\section{Masses of FOF halos}
\label{sec:masses}

%
\begin{figure}[t]
\includegraphics[scale=0.95]{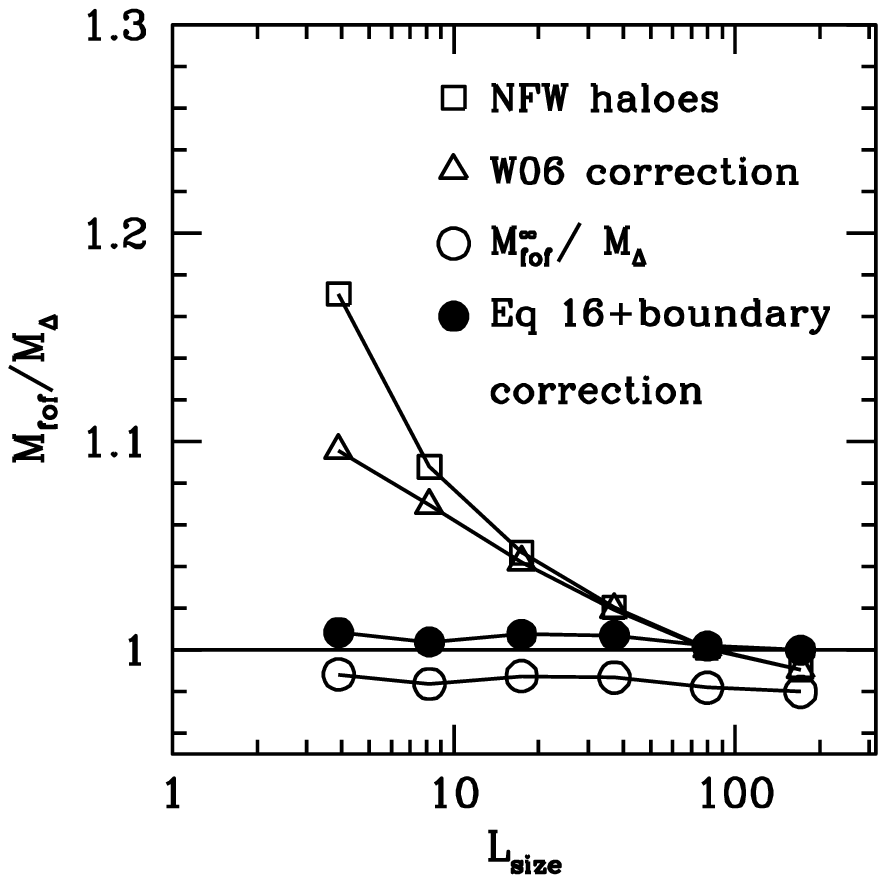}
\caption{
    The fraction, $M_{\rm fof}/M_{\Delta}$, where $M_{\rm fof}$ is the
    halo mass selected by the FOF algorithm and $M_{\Delta}$ is the
    mass within the overdensity given by Eq.~\ref{eq:od} as a function
    of the resolution with which the halo is sampled. Squares show the
    fraction obtained by running FOF on our simulated NFW halos.
    Triangles show the fraction after the FOF masses were corrected by
    the formula given by \citet{warren_etal06}. Open circles show the
    fraction predicted by Eq.~\ref{eq:mcorr} and it corresponds to the
    fraction if the FOF algorithm was run on a halo with infinite
    resolution. Finally, filled circles show the fraction after
    correcting the open circles for the boundary profile of halos
    selected by FOF.
  }
\label{figwarren}
\end{figure}
\begin{figure}[t]
\includegraphics[scale=0.95]{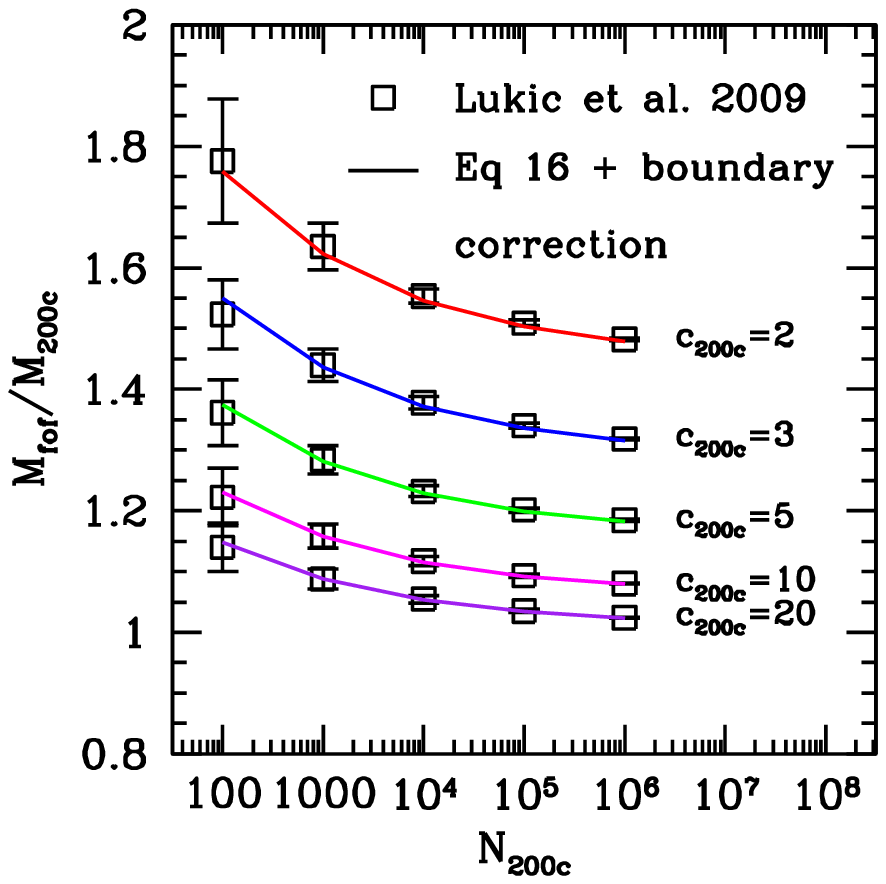}
\caption{
 The mass of halos selected by the FOF algorithm with $b=0.2$ relative to
 the mass within overdensity of 200 times the critical density of the Universe for
 halos with different concentration, $c_{200c}\equiv R_{200c}/r_s$. 
Results of Monte Carlo realizations of spherical NFW halos by 
 \citet{lukic_etal09} are shown by squares, while predictions of our
 model for each concentration given by eq.~\ref{eq:mcorr} and after
 applying the additional FOF boundary correction are shown by solid lines.
}
\label{figlukic}
\end{figure}
\subsection{Masses of the idealized FOF halos in the context of percolation theory}

Using Monte Carlo simulations of isothermal halos with varying
numerical resolution, \citet{warren_etal06} were the first to
demonstrate that the mass of halos selected by the FOF algorithm
depends upon the resolution with which the halo is sampled.  They
found that at lower resolutions the FOF algorithm assigns
systematically larger masses to halos. They devised an empirical
formula to correct the effects of such systematic bias on the halo
mass function. More recently,
\citet{lukic_etal09} carried out Monte Carlo simulations of NFW halos
and found a qualitatively similar effect \citep[see
also][]{bhattacharya_etal10}. They also devised an empirical formula
to correct for the resolution-dependent mass bias for the specific
case of $b=0.2$ and idealized spherical NFW halos that they
studied. \citet{lukic_etal09} showed that this correction depends not
only on the number of particles but also upon the concentration of the
halo.

As can be seen from Figure~\ref{fig1}, our experiments also reveal a
qualitatively similar effect. The boundary identified by the FOF
algorithm significantly widens with decreasing number of halo
particles.  Therefore, the mass selected by the FOF algorithm also
increases with decreasing number of particles. In
Figure~\ref{figwarren}, we show the mass of the halo identified by FOF
for each of our spherical Monte Carlo halos normalized by $M_\Delta$,
the mass expected within the overdensity predicted by using
Eq.~\ref{eq:od}. We plot this quantity as a function of $L_{\rm size}$
given by
\begin{equation}
L_{\rm size} = \frac{2 R_{\Delta}}{b\bar{l}}=\frac{2}{b}\left(\frac{3
N_{\Delta}}{4 \pi \Delta}\right)^{1/3} \,.
\end{equation}
Note that by definition $L_{\rm size}$ approximately corresponds to the
inverse of the fractional accuracy with which a halo boundary can
ever be identified by the FOF algorithm and it depends upon the
resolution of the halo via $N_{\Delta}$. As described in the appendices,
$L_{\rm size}$ is thus the
appropriate parameter to use from the standpoint of percolation theory
to parameterize the dependence of FOF mass for a given halo on the
numerical resolution. 

Figure~\ref{figwarren} shows that FOF mass can be systematically
biased high by $\approx 10-20\%$ for $L_{\rm size}\lesssim 10$. Most
of the modern state-of-the-art simulations are in this regime. For
example, the Bolshoi and MultiDark simulations used in the previous
section, followed evolution of $2048^3\approx 8.59\times 10^9$
particles in boxes of $250h^{-1}$~Mpc and $1000h^{-1}$~Mpc,
respectively. For $b=0.2$, these simulations have $b\bar{l}$ of
$\approx 24.4h^{-1}$~kpc and $\approx 97h^{-1}$~kpc,
respectively. Thus, $L_{\rm size}\leq 10$ corresponds to halos with
virial radii $R_{\Delta}\leq 122h^{-1}$~kpc and $R_{\Delta}\leq
488h^{-1}$~kpc, respectively, both well within the range of halos
resolved by these simulations. A wider range of masses would be
affected for lower resolution simulations. Dependence of $L_{\rm
size}$ on the number of particles in a halo for the choice of $b=0.2$
and typical halo concentration is presented in Figure~\ref{fig10} in
the Appendix, which shows that $L_{\rm size}\lesssim 10$ for
$N_{\Delta}\lesssim 10^4$.

In the Appendix, we show that the extra mass identified by the FOF algorithm at
a given resolution (i.e., a given $L_{\rm size}$) can be 
accurately corrected by the following formula motivated by percolation theory:
\begin{equation}
M^{\infty}_{\rm fof} = M_{\rm fof}\, \left( 1 + 0.22\,\alpha
\,L_{\rm size}^{-1/\nu}\,\left|\frac{\partial\ln M_{\Delta}}{\partial
p}\right|
\right)^{-1}\,.
\label{eq:mcorr}
\end{equation}
Here, $M_{\rm fof}^{\infty}$ denotes the mass of the halo that FOF
would identify at infinite resolution, $\nu$ is a critical exponent
from percolation theory and is $\approx 1.33$ in our case
(see the Appendix for details), $\alpha$ denotes the logarithmic slope
of the halo density profile at the percolation theory predicted
boundary, $R_\Delta$. For an NFW density profile, $\alpha$ is given by
\begin{equation}
\alpha = 1 + \frac{2\,c_{\Delta}}{1+c_\Delta} \,.
\end{equation}
The probability $p(r)$ (see Appendix for the connection to percolation
theory) at a given radius depends upon the number density of particles at that
radius, $n(r)$, via
\begin{equation}
p(r) = 1 - \exp\left\{ -\frac{\pi}{6} (b \bar{l})^3 n(r) \right\}\,,
\end{equation}
and $\partial \ln M_{\Delta}/ \partial p$ denotes the derivative of the logarithm of the
mass with respect to $p$ at the percolation threshold predicted
boundary, $R_{\Delta}$. Larger values of $L_{\rm size}$ correspond to
higher resolution and the mass measured by the FOF algorithm tends to
$M_{\rm fof}^\infty$ asymptotically. Note that our correction formula
depends upon the number of halo particles, $N_{\Delta}$, the linking length
parameter $b$, and the concentration parameter, $c_{\Delta}$.

The circles in Figure~\ref{figwarren} show the result of this
correction. The figure shows that the mass corrected by this formula
is independent of $L_{\rm size}$. The triangles, on the other hand,
show the empirical correction of \citet{warren_etal06}, which clearly
fails to correct the effect fully. This is not surprising as this
formula was devised to correct resolution bias {\it in the halo mass
function}, rather than mass of individual idealized NFW halos. As we
show below, other resolution effects affect masses of real CDM halos
and thereby the halo mass function.  The presented exercise simply
indicates that the formula of \citet{warren_etal06} does not describe
the mass bias of idealized halos considered here.

Also note that even at infinite resolution the FOF algorithm selects a
mass which is smaller than $M_{\Delta}$ by $\approx 2\%$. This is
because the boundary of FOF halos is not a step function even at
infinite resolution (see Fig.~\ref{fig1}).  We defer detailed
discussion of this effect to the Appendix and show that this small
additional correction can also be calculated from 
percolation theory. The bold circles in Fig.~\ref{figwarren} show the
result of correcting the masses taking into account this additional
small effect. As the figure shows, the full correction brings the value of the FOF halo masses
in good agreement with the true mass $M_{\Delta}$.

Figure~\ref{figlukic} shows the results of the Monte Carlo
realizations of spherical NFW halos of differing concentrations carried out by
\citet[][shown by squares]{lukic_etal09} and predictions of our model (shown by solid lines). 
These authors applied the FOF algorithm with $b=0.2$ to identify
halos from the realizations and showed that FOF mass of halos
depends on concentration of their density
distribution. \citet{lukic_etal09} defined both the reference halo
mass, $M_{200c}$, and concentration, $c_{200c}$, relative to the
radius, $R_{200c}$, enclosing overdensity of 200 times the critical
density of the universe. They found that FOF mass is generally
significantly different than $M_{200c}$ and the difference depends on
$c_{200c}$ and the number of particles in a halo (effect similar to
that discussed above).

We show our percolation theory-motivated prediction for the ratio of
the FOF halo masses to $M_{200c}$ calculated by using
Eq.~\ref{eq:mcorr} and after applying the correction for the boundary
of the halo as solid lines in Figure~\ref{figlukic}. The
prediction is in excellent agreement with the results
of \citet{lukic_etal09} and it accurately captures the dependence of
$M_{\rm fof}/M_{200c}$ ratio on the concentration and particle number
found by \citet{lukic_etal09}. We would like to note that the
correction formula presented \citet{lukic_etal09} is a numerical fit
to their results and is only valid for the linking length parameter,
$b=0.2$ for which they calibrate their correction. The correction
based on equation~\ref{eq:mcorr} is valid for different values of $b$,
concentrations, and values of the numerical resolution ($L_{\rm
size}$).

In the Appendix, we also test our correction against simulated halos
with varying slopes of the number density profile and show that it
works remarkably well for different slopes. We also show that we are
able to explain the empirical results for isothermal
halos\footnote{We note that the empirical formula given
by \citet{warren_etal06} does not explain the results of their
isothermal halos.} found by \citet{warren_etal06}.

Given that the density of CDM halos decreases rapidly near the outer
virialized regions, an overestimate of mass for small $L_{\rm size}$ and
$N_{\Delta}$ corresponds to an {\em underestimate} of the enclosed
overdensities of FOF halos.  This underestimate can be seen in the
form of downturn of overdensity for halos from $\Lambda$CDM
simulations observed in Figures~\ref{fig5} and \ref{fig6}. For a fixed
mass and fixed value of $b$, the Bolshoi simulation has a larger value
of $L_{\rm size}$ than the MultiDark simulation. This explains why the
downturn occurs at lower halo masses for the Bolshoi than for the
MultiDark simulation. It is also clear from Eq.~\ref{eq:mcorr}, that
$L_{\rm size} \propto b^{-1}$, and therefore the downturn in
overdensity shifts to smaller masses for decreasing values of $b$.

\subsection{Resolution dependence of the FOF mass for real $\Lambda$CDM halos}
\label{sec:corr}
%
\begin{figure*}[t]
\centering
\includegraphics[scale=0.9]{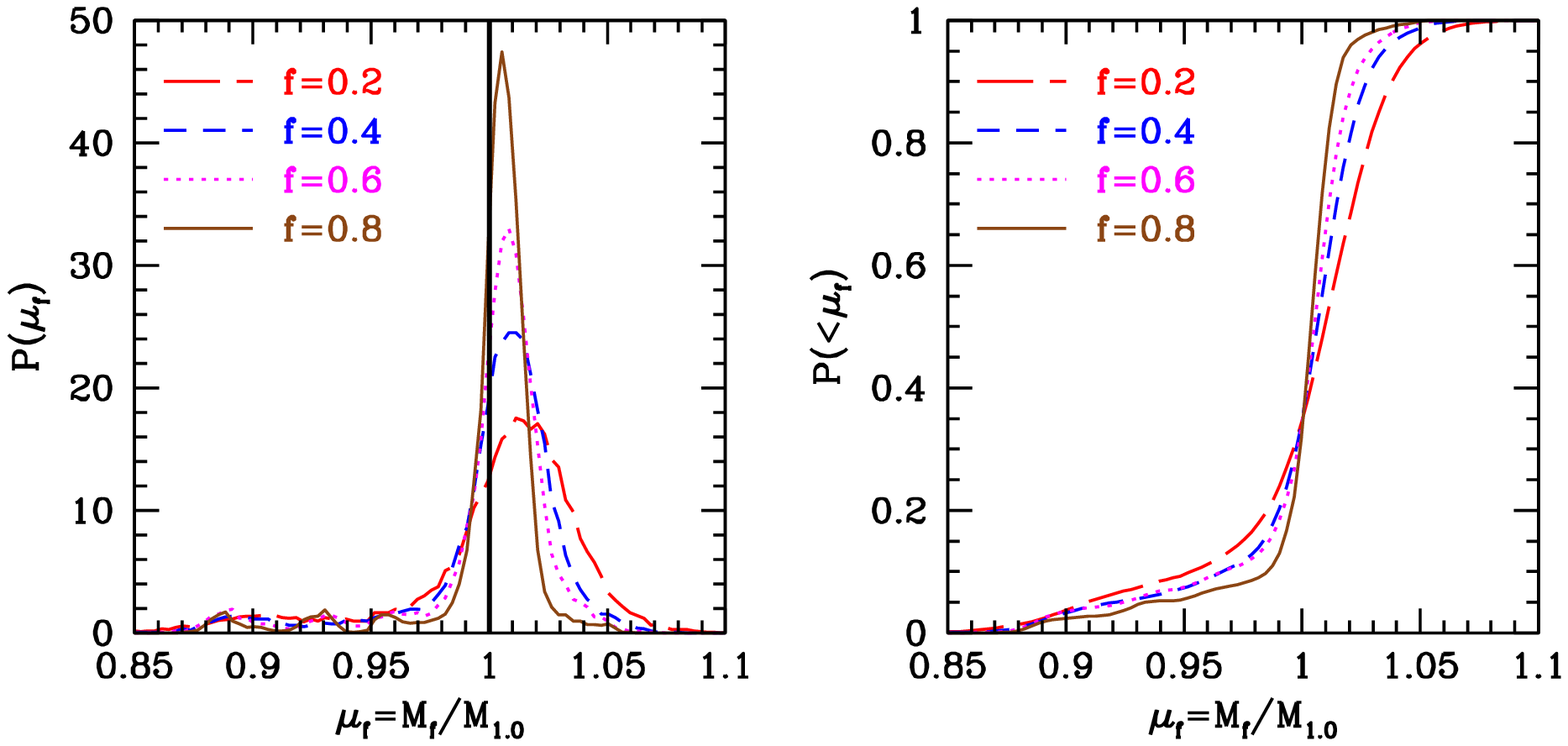}
\caption{
    The left hand panel shows the probability distribution of the
    ratio $\mu_{\rm f}$ of the mass selected by the FOF algorithm when applied to the $1000$
    subsamples of a fraction $f$ of the particles around the $25$ most
    massive halos to the mass of the halo selected by the FOF algorithm with
    $f=1.0$. The right hand panel shows the cumulative probability
    $P(<\mu_{\rm f})$. Different line types are used to indicate the result
    obtained for different values of $f$.
  }
\label{fig:withsub}
\end{figure*}
\begin{figure*}[t]
\centering
\includegraphics[scale=0.85]{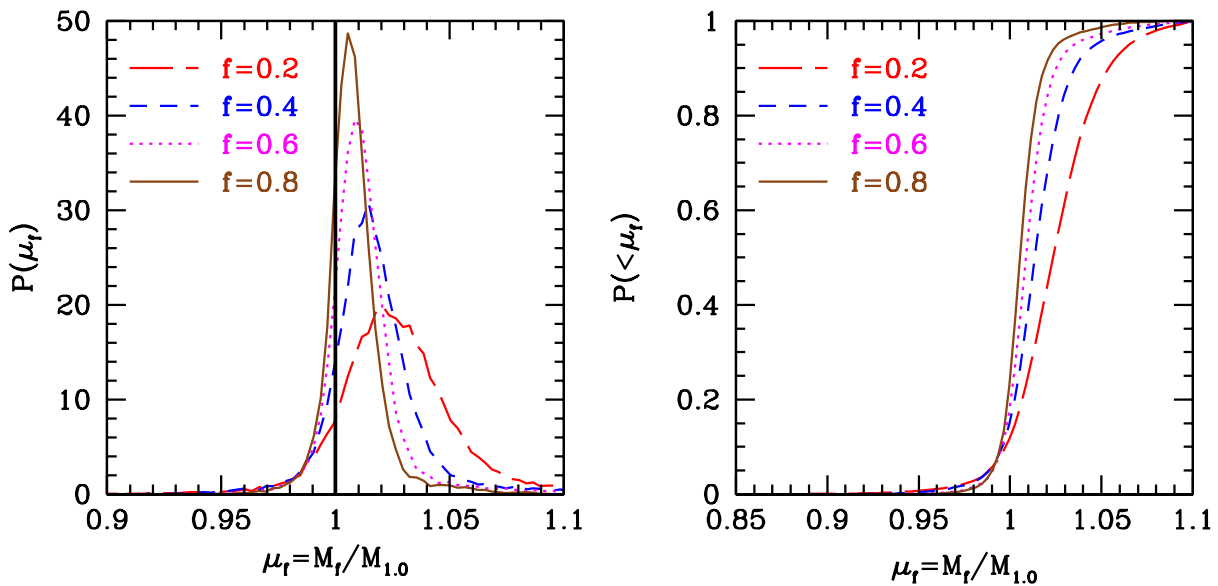}
\caption{
    Same as Fig.~\ref{fig:withsub}, except when the angular
    coordinates of the particles around the center of the FOF halo are
    shuffled to disperse substructure (see text for details).
  }
\label{fig:nosub}
\end{figure*}
\begin{figure*}[t]
\centering
\includegraphics[scale=0.85]{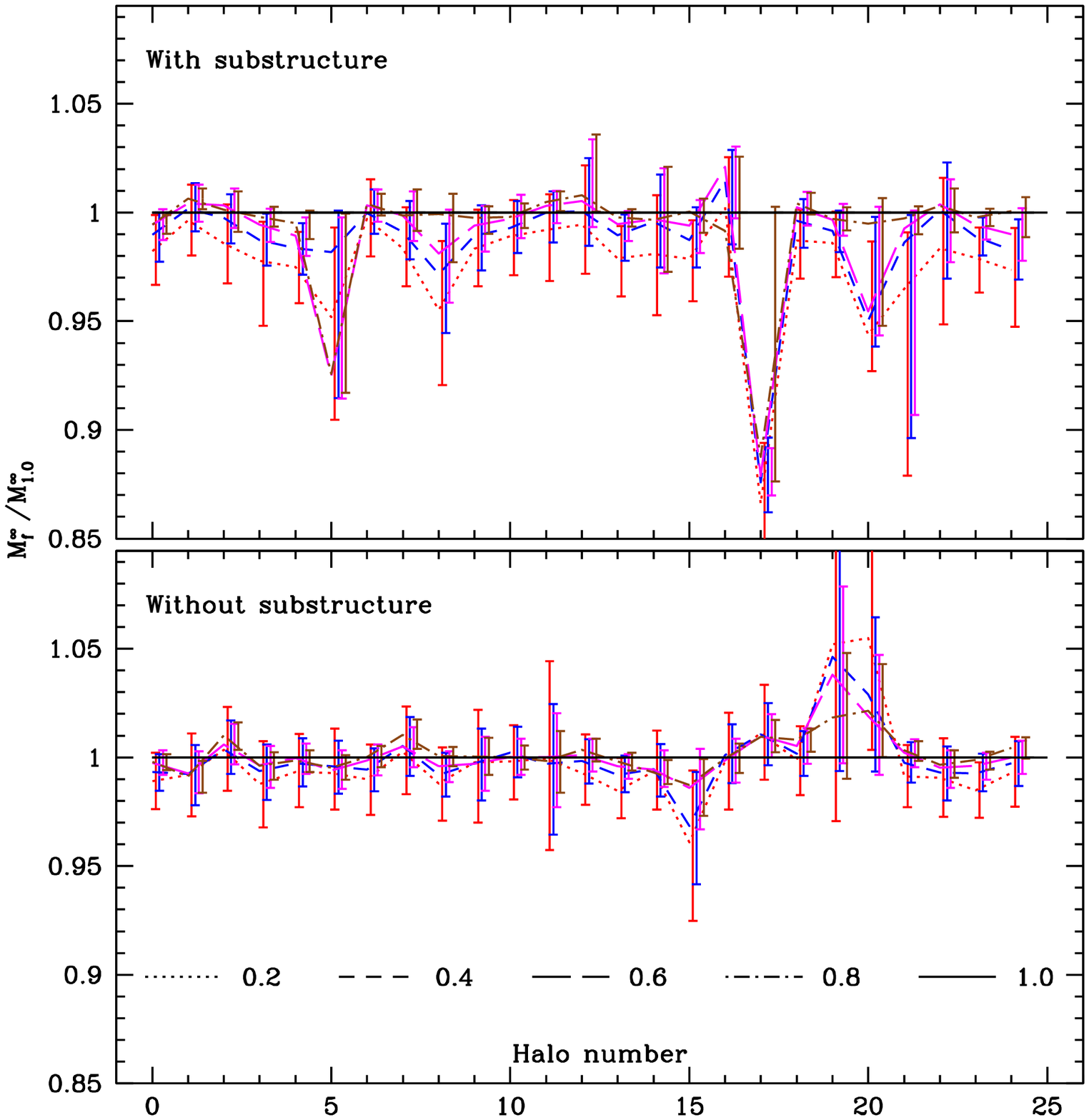}
\caption{
    The ratio $M^{\infty}_{\rm f}/M^{\infty}_{1.0}$ for
    the 25 most massive halos selected from the simulation. Here,
    $M^{\infty}_{\rm f}$ denotes the median of the distribution
    of masses selected by the FOF algorithm when run on a fraction $f$ of the
    particles after correcting for the finite size effect using
    eq.\ref{eq:mcorr}. The top panel shows the result of the real
    halos, while the bottom panel shows the results when the angular
    coordinates of the particles around the center of the FOF halo are
    shuffled to disperse substructure (see text for details).  As
    indicated in the legend, different line types are used to indicate
    different values of the fraction $f$. The errorbars are used to
    indicate the 16 and 84 percentile of the distribution. The
    errorbars for different values of $f$ are shifted in the $x$
    direction for clarity.
  }
\label{fig:testcorr}
\end{figure*}
In the previous subsection, we showed that the mass of halos selected by
the FOF algorithm depends upon $L_{\rm size}$. The mass $M$ selected
by FOF at finite $L_{\rm size}$ can be larger than $M^\infty$ by as
much as 5 - 20\% for small values of $L_{\rm size}$. This effect, if
not corrected for, can potentially introduce systematic errors in the
determination of the mass function using halos selected by
FOF. We have also shown that the percolation theory
motivated formula given by eq.~\ref{eq:mcorr} is able to correct this
dependence of the mass on $L_{\rm size}$ for spherical NFW halos (or
for spherical halos with a power law density profile). Real halos,
however, are not spherical and contain substructure.  In this section,
we therefore test the correction formula derived for idealized halos
against undersampled versions of real halos selected from
cosmological simulations.

For this purpose, we make use of the L1000W simulation of size $L_{\rm
B}=1\gpch$, described in detail in \citet{tinker_etal08}. The
simulation follows the evolution of dark matter particles in a
$\Lambda$CDM cosmology with parameters that are slightly different
from the Bolshoi and the MultiDark simulation: the matter density and
the baryon density in units of the critical density, $\Omega_{\rm
m}=0.27$ and $\Omega_{\rm b}=0.044$, the Hubble constant $h=H_0/(100
\kmsmpc)=0.70$, the rms amplitude of linear fluctuations in spheres of
radius $8h^{-1}$~Mpc, $\sigma_8=0.79$ and the power law slope of the
initial power spectrum, $n_s=0.95$. We run the FOF algorithm with a
linking length parameter $b=0.2$ on the redshift zero snapshot of the
simulation. For the purpose of our tests, we focus our attention to the
25 most massive halos selected by FOF.

We selected all particles within a radius $R_{\rm max}=10 \mpch$ of
the center of mass of each of these halos. We have verified that all
the particles of each halo selected by FOF lie well within $R_{\rm
max}$. We created $1000$ subsamples each of particles around every
halo by using only a fraction $f \in \{0.2,0.4,0.6,0.8\}$ of the
particles. We then run FOF on each of these subsamples using a linking
length parameter $b=0.2\,f^{-1/3}$. We use the symbol $\mu_f$ to denote
the ratio of the mass selected by FOF when run on a subsample with a
fraction $f$ of the original particles to the mass of the FOF halo
when using all the particles. 

In the left hand panel of Fig.~\ref{fig:withsub}, we show the
distribution of $\mu_f$ for different values of $f$ using different line
types. Note that the peak of the distribution shifts towards larger
values of $\mu_f$ for smaller values of $f$. This is qualitatively
similar to the behavior of FOF discussed in \S~\ref{sec:masses}.
However, we also notice that the distribution of $\mu_f$ has a
significant tail towards smaller values of $\mu_f$. In roughly one third
of the cases (9 out of 25), the FOF algorithm often fails to bridge a
structure in the outer parts of the halo with the main halo. The effect appears less severe because we have
plotted the combined distribution of $\mu_f$ values for the 25 halos.
However, in the case of halos for which bridging is an issue, the
distribution of $\mu_f$ clearly shows a bimodal distribution. 

The right hand panel of fig.~\ref{fig:withsub} shows the cumulative
distribution of $\mu_f$. Note that smaller values of $f$ have a slightly
larger tendency to avoid bridging. This counteracts the tendency to
select larger masses at smaller values of $f$. If we assign a mass for
each halo for a given value of $f$ as the average of the FOF mass over
the $1000$ subsamples, we often find that this average FOF mass
increases as $f$ increases contrary to our idealized NFW halos.
Clearly using the average is sensitive to the tails of the
distribution. Therefore, we used the median of the FOF masses of the
$1000$ subsamples to test our correction formula.

We denote the median mass selected by the FOF algorithm when run on a fraction $f$
of the particles by $M_{\rm f}$ and the median mass after
correcting for the finite size effect using eq.\ref{eq:mcorr} by
$M^\infty_{\rm f}$. The top panel of Figure~\ref{fig:testcorr} shows
the ratio of $M^\infty_{\rm f}/M^\infty_{1.0}$ for the
25 most massive halos. Our correction formula, which worked extremely
well for the idealized spherical NFW halos, seems to systematically
overcorrect for the finite size effect for small values of $f$ by
$\approx 3 - 5\%$. 

The two plausible causes for this behavior are: (i) the
non-sphericity of real halos, and (ii) the presence of substructure in
real halos. We carried out another set of Monte-Carlo simulations of
idealized triaxial halos where the number density of particles is
given by a NFW-like profile with the radius $r$ replaced by $\zeta$ such
that
\begin{equation}
\zeta^2 = \frac{x^2}{a^2}+\frac{y^2}{b^2}+\frac{z^2}{c^2}
\end{equation}
We used values of $a/c=0.6$ and $b/c=0.8$, typical for halos found in
numerical simulations of dark matter. We have verified that the
correction formula given by eq.\ref{eq:mcorr} works perfectly well
even if our triaxial halos are incorrectly assumed to be spherical.
Our use of the spherically averaged number density distribution to
determine the correction does not introduce any systematic errors. We
also experimented with particles whose number density distribution
follows a power law in radius and found an identical result.

To investigate the effects of substructure, we carried out the
following test. We first obtained the SPH estimate of the density at
the location of all particles in each of the halos using 128 nearest
neighbor particles. We used the position of the particle with the
largest density as the center of the halo. We then randomly reassigned
the angular coordinates of each of the particles within a 10 $\mpch$
sphere with respect to the center of the halo. In this manner, we were
able to disperse the substructure over a wider range of angular
coordinates while still preserving the radially averaged density
profile. We then repeated our exercise of running FOF on subsampled
versions of this set of particles. 

We show the results of this exercise in Figure~\ref{fig:nosub}, which
shows the distribution of values of $\mu_f$ thus obtained. In contrast
to Figure~\ref{fig:withsub}, the distribution of $\mu_f$ is much more
symmetric with no significant presence of tails. The peak of the
distribution occurs at larger values of $\mu_f$ as $f$ is decreased.
The lower panel of Figure~\ref{fig:testcorr} shows the ratio
$M^\infty_{\rm f}/M^\infty_{1.0}$ for halos where the substructure has
been dispersed. Contrary to the results in the top panel, in this case
our correction formula corrects masses accurately.  This shows that
failure of the correction formulae derived for idealized halos is due
to substructure present in real CDM halos simulated with sufficiently
high resolution.

The results of this exercise show that the masses selected by FOF for
realistic halos can not be corrected for finite size effects in a
straightforward manner. {\it Although percolation-motivated correction
formula we derived for halos without substructure (eq.~\ref{eq:mcorr}) is highly accurate,
it cannot be blindly applied to correct halo masses selected by the
FOF algorithm.} Substructure introduces strong resolution-dependent
effects. The amount of substructure depends on resolution of
simulations in a non-trivial way and will vary for halos of different
mass within a simulation. It will also vary with redshift for a given
halo mass. This indicates that any empirical formula designed to
correct masses of halo mass function for resolution effects will also
depend in a non-trivial way on resolution, cosmology, and redshift. We
thus caution against the use of empirical formulae that depend just
upon the number of particles in a halo calibrated for a single
cosmology and redshift, as these will likely be inaccurate for other
cosmologies and redshifts.

\section{Discussion and conclusions}
\label{sec:conc}

In this paper we have explored properties of halos identified by the
FOF algorithm focusing on the halo boundary. Using idealized Monte
Carlo realizations of spherical NFW halos we showed that boundary of
the FOF halos spans a range of local overdensities and is inherently
``fuzzy.'' The fuzziness of the boundary increases with decreasing
number of halo particles. We demonstrate that these results can be
interpreted in terms of the percolation theory, which we discuss in detail in the Appendix. The value of
characteristic local overdensity within FOF boundary derived from our
Monte Carlo realizations and predicted by percolation theory is given
by (eq.~\ref{eq:dfof}): $\delta_{\rm fof}=0.6529b^{-3}-1$, which gives
$\delta_{\rm fof}=80.61$ for the commonly used value of $b=0.2$. This
is significantly larger than the local overdensity of $\approx 60$
usually assumed for this value of linking length. Correspondingly, the
enclosed overdensity of typical FOF halos is significantly larger than
$180$ and ranges from $\sim 250$ to $\sim 600$. Specific value of the
enclosed overdensity is determined by the concentration of halo
(density distribution) and therefore depends on cosmology, halo mass,
and redshift. We predict this dependence using a simple analytic model
based on NFW density profile and show that this model reproduces
results of cosmological simulations of $\Lambda$CDM cosmology at
different halo masses, redshifts, and values of the linking length
$b$.

For a given linking length $b$, the range of overdensities (i.e., the
fuzziness) in the boundary of FOF halos increases with decreasing
number of halo particles due to changing properties of percolation for
smaller values of parameter $L_{\rm size}\equiv
2R_{\Delta}/(b\bar{l})$, where $R_{\Delta}$ is the effective radius of
the FOF boundary. For a given simulation, this results in a systematic
and increasing overestimate of the FOF mass with decreasing halo
mass. This effect has been found empirically by \citet{warren_etal06}
and \cite{lukic_etal09}. 

We demonstrate how it can be understood qualitatively on the basis of
percolation theory. We also present an accurate formula for correcting
this systematic FOF mass bias for idealized halos without
substructure. This formula is accurate for different values of linking
lengths $b$, halo concentrations, and values of parameter $L_{\rm
size}$. We note, however, that this accurate correction requires
knowledge of the halo concentration-mass relation, which itself would
need to be accurately calibrated for different cosmologies. Moreover,
as we demonstrated in \S~\ref{sec:corr}, substructure in real halos
introduces additional substantial resolution-dependent biases into
masses of FOF halos.  Given that amount of substructure depends on
resolution of simulations and simulation cosmology and redshift in a
non-trivial way, any empirical mass correction formula should also
depend in a non-trivial way on resolution, cosmology, and redshift.

The concentration and non-trivial resolution dependence of enclosed overdensities
and masses of the FOF halos make it difficult to interpret their raw
mass function and its universality physically in terms of an underlying
model of nonlinear collapse. For instance, as we note
in \S~\ref{sec:mf}, concentration dependence of FOF overdensity is
likely behind smaller deviations of the FOF halo mass function from
universality, as some of the real redshift evolution of the halo mass
function is partially cancelled by redshift evolution of halo
concentrations. Although such partial cancellation may work for a
single $\Lambda$CDM cosmology, it will not work in general as halo
concentrations do depend on cosmological parameters. All this also
makes it more complicated to connect FOF halo masses to observational
estimates of masses, which are typically made within spherical
apertures enclosing a fixed (and fairly high) overdensity,
with concentration of density profile not known a priori.

Neverthless, results of \citet{courtin_etal10} do indicate that
universality of the halo mass function can be improved if cosmology
dependence of non-linear virialization is taken into account properly
in the definition of halo mass. In \S~\ref{sec:mf}, we show that their
empirical findings can be understood better in terms of our results
and model. Further exploration of this issue is definitely warranted.
Overall, even though interpretation of FOF halo statistics is more
complicated in light of our results, improved understanding of the FOF
identified halos makes any interpretation more robust.

Our results should be also useful in constructing mock catalogs of
galaxies based on FOF halo catalogs. To reproduce galaxy clustering
properly this procedure requires good knowledge of internal
overdensity of identified halos. Model and percolation theory results
presented in this paper can be used to accurately estimate this
overdensity even for halos with small numbers of particles.


\section*{Acknowledgments}

SM and AVK are supported by the Kavli Institute for Cosmological
Physics at the University of Chicago through the NSF grant PHY-0551142
and an endowment from the Kavli Foundation.  AVK is also supported by
the NSF grants AST-0507596 and AST-0708154. SM and AVK are grateful to
the members of structure formation group at the University of Chicago
for many useful discussions. AVK is also grateful to Aleksander
Kravtsov for patience during completion of this paper. SG acknowledges
support of DAAD through the PPP program.  The MultiDark simulation
used in this paper were performed and analyzed at the NAS Ames
Research Center. We thank A. Klypin (NMSU) and J. Primack (UCSC) for
making these simulations available to us. These simulations will be
available to the community via the MULTIDARK database
{\tt http://www.multidark.org/MultiDark/} as part of the activities of the
German Astrophysical Virtual Observatory.
We would like thank Zarija Lukic (LANL) for providing data from their paper in
electronic form and we are grateful to him and Frank van den Bosch for
comments on an early draft of this paper. SM would also like to thank
Robert Ziff for interesting discussions on percolation theory. This
work made extensive use of the NASA Astrophysics Data System and {\tt
arXiv.org} preprint server.


\appendix
\section{Brief review of the relevant aspects of percolation theory}
\label{sec:perco}

Consider a point process that generates a set of points on an
$N$-dimensional manifold. Percolation theory deals with the statistics
of clusters (or groups of friends in FOF terminology) formed by grouping
together {\it neighboring} points on the manifold. Traditionally, the
percolation problem is defined on a lattice  where the
occupation of each lattice cell is determined by a random process
\citep{stauffer}. However, the continuum percolation (Swiss-cheese)
model is more relevant to our discussion of the FOF algorithm
\citep{roberts68, domb72, lorenz_ziff01}. In this appendix, we briefly
describe this model and how the profile of the boundary of a FOF halo
can be understood in more detail.

The Swiss-cheese percolation model considers a set of spheres of equal
radius, $R$, whose centers are distributed by a random Poisson process
with a {\it constant} average number density $n(\vecb{x})$ in a
$L\times L\times L$ volume, where $L\gg R$. The spheres can be thought
of as spheres carved in a slab of cheese, from which the model derives
its name. Groups of overlapping spheres form clusters of varying
sizes. The largest cluster that forms in the system is of particular
importance, and for a fixed value of $R$, its size depends upon the
average number density of spheres in the system. As the number density
of spheres is increased, the size of the largest cluster increases
until at a critical number density the largest cluster size becomes
$\approx L$. This event is called percolation, the smallest number
density at which it happens is called the critical percolation
threshold and the corresponding cluster is called the infinite
cluster. The critical density, $n_c$ in units of $1/(2\,R)^3$ is a
universal constant and has been accurately measured by extensive
Monte-Carlo simulations: $n_c=0.652960\pm0.000005$
\citep{lorenz_ziff01}. 

The linking length of the FOF algorithm, $b\bar{l}$, corresponds to
the diameter $2\,R$ of the spheres in the Swiss-cheese percolation
model. The centers of overlapping spheres correspond to ``friend''
particles in the FOF algorithm as the distance between the centers is
less than the linking length. In the FOF language, the critical
density threshold is therefore $n_{\rm crit}\equiv
n_c/(2R)^3=n_cb^{-3}\bar{l}^{-3}$, which corresponds to an
overdensity of $\delta=n_{\rm crit}/\bar{n}-1=n_cb^{-3}-1$.

For the Swiss-cheese model, the probability for any given point
$\vecb{x}$ in the $L\times L\times L$ volume to belong to a non-zero
number of spheres is given by
\begin{equation}
p(\vecb{x})=1-\exp\left\{-\frac{4}{3} \pi R^3
n(\vecb{x})\right\}=1-\exp\left\{-\frac{1}{6} \pi (2 R)^3
n(\vecb{x})\right\}\,.
\end{equation}
It is conventional to define the percolation problem in terms of this
probability instead of the number density $n(\vecb{x})$, in which case the critical threshold for percolation $p_c$ is
related to $n_c$ via
\begin{equation}
p_c = 1-\exp\left(-\frac{\pi}{6} n_c\right)\,.
\end{equation}
Close to the percolation threshold, the probability that any point
$\vecb{x}$ belongs to the infinite cluster, $\pinf$, also called the
strength of the infinite cluster, follows the scaling relation
\begin{equation}
\pinf \approx (p-p_c)^{\beta} \,,
\label{eq:pinf}
\end{equation}
where $\beta$ is a constant which depends upon the dimensionality of
the problem. Only few problems in percolation have exact analytical
solutions. Hence, the constant $\beta$ has to be determined by
Monte-Carlo experiments and it has been found to approximately equal
to $0.42$ for percolation in three dimensions. \citep[see,
e.g.,][]{stauffer}. Another quantity of interest is the correlation or
the connectivity length, denoted by $\xi$, and defined as the average
distance between two points that belong to the same cluster. As $p$
approaches $p_c$, $\xi$ follows the scaling relation given by
\begin{equation}
\xi \propto |p-p_c|^{-\nu}\,
\label{xi_scale}
\end{equation}
where the constant $\nu$ again depends upon the dimensionality of the
problem and is approximately equal to $0.88$ in three dimensions and
$4/3$ in two dimensions. 

How do these basics of the percolation theory relate to the
halos identified by the FOF algorithm? In the context of the Monte Carlo
realizations of
spherical NFW halos considered in \S~\ref{sec:od}, the particle
distribution of a given realization is a set of points distributed in
a spherical volume of radius $2R_{180}$. The FOF algorithm with
linking length $b$ applied to these points treats particles as a set
of spheres of radius $R=b\bar{l}/2$. Those particles whose spheres overlap
are considered friends. The difference from a simple uniform
density example considered above is that our halos have non-uniform
density distribution. Thus, instead of considering percolation in a
uniform distribution for different particle number densities, we are
considering percolation as we decrease the number density of particles as
a function of increasing radius. For a given $b$, there will be a
certain radius at which the critical number density for percolation,
$n_c$ (and corresponding probability $p_c$) is reached. Particles
around this radius will have a high probability $\pinf$ to be a part
of the infinite cluster -- i.e., to be joined into FOF halo. It is
these particles that form the boundary of an FOF halo.  Below we
consider the properties of this boundary in the context of the
percolation theory.

\section{Detailed analysis of the FOF boundary of NFW halos}
\label{sec:bdry}

\begin{figure}[t]
\centering
\includegraphics[scale=0.8]{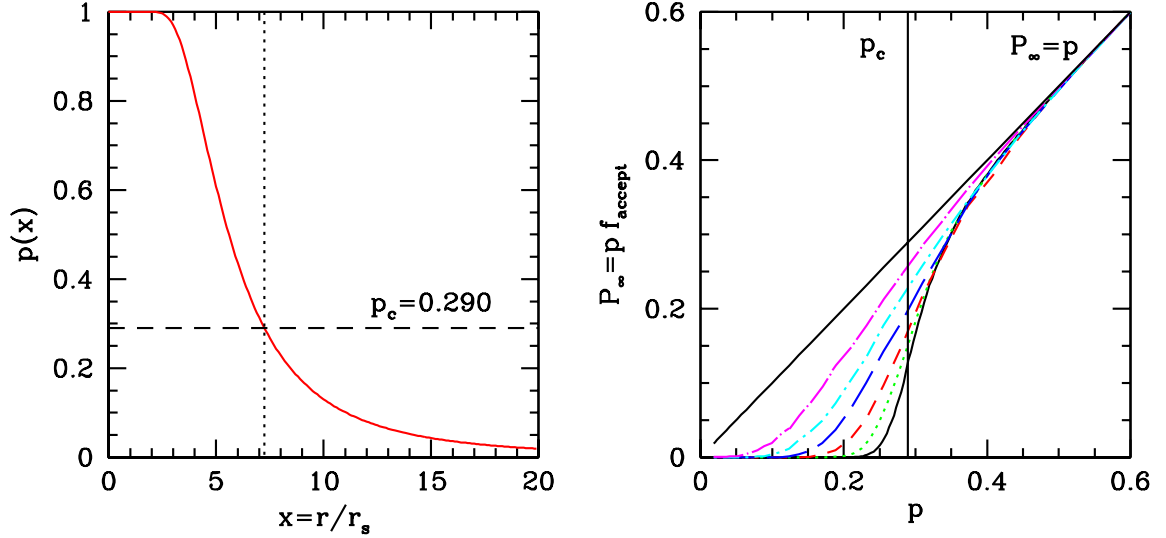}
\caption{ 
    The probability $p$ as a function of the radius (left panel) and
    probability to be a part of an infinite cluster, $\pinf$, as a
    function of $p$ (right panel) for the Monte Carlo realizations of
    spherical NFW halos ($c=10$) analyzed in \S~\ref{sec:od}. In the
    left panel the critical threshold for percolation $p_c$ is shown
    with the horizontal dashed line. In the right panel $p_c$ is shown
    by the solid vertical line; different line types correspond to
    halo realizations with different numbers of particles, with line
    types and colors corresponding to the same halos as in
    Figure~\ref{fig1} (from left to right lines correspond to $N_{180}$
    from $100$ to $10^7$ particles.  }
\label{fig8}
\end{figure}

In the left panel of Figure~\ref{fig8}, we show the probability $p$ for
a point to be within a distance $b\bar{l}/2$ from any particle as a
function of its position $x=r/r_s$ for the Monte Carlo realizations of
spherical NFW halos analyzed in \S~\ref{sec:od}. In percolation
theory, for point distributions with non-uniform density the infinite
cluster is defined as the cluster connected to spheres that lie in
the region where the probability $p \rightarrow 1$. In our case, this is
equivalent to the group that consists of particles at the center of
the halo and is the largest group found by the FOF algorithm.

\begin{figure}[t]
\centering
\includegraphics[scale=0.8]{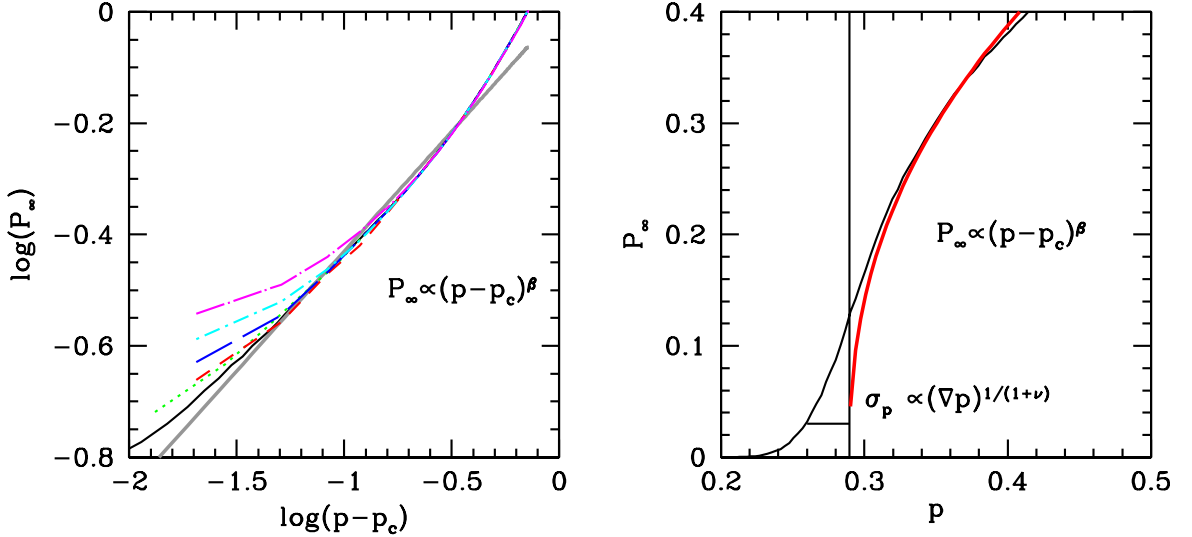}
\caption{ 
   The left hand panel shows the strength of the infinite cluster,
   $\pinf$, as a function of $p-p_c$ for our Monte Carlo realizations
   of spherical NFW halos. Different line types correspond to halos
   generated with varying numbers of particles. Line types and colors
   correspond to the same halos as in Figure~\ref{fig1}.  The right
   hand panel shows the strength as a function of $p$, for the highest
   resolution halo. The solid red line shows prediction of the
   percolation theory for a uniform distribution of particles.}
\label{fig9}
\end{figure}
\begin{figure}[t]
\centering
\includegraphics[scale=1.0]{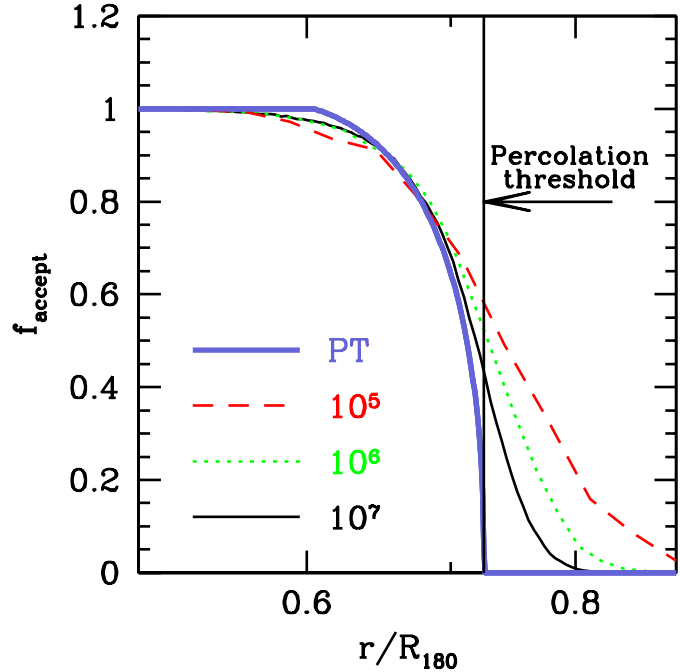}
\caption{ 
   The fraction of particles that are joined by the FOF algorithm
   (with $b=0.2$) into the main halo as a function of the radius in
   units of $R_{180}$ for our Monte Carlo realizations of spherical
   NFW halos. Bold solid line shows the percolation theory prediction
   for uniform particle density, which can be compared to the results
   of our simulations shown with lines of different style and
   color. Number of particles in each halo realization is indicated in
   the legend.  }
\label{fig11}
\end{figure}
\begin{figure}[t]
\centering
\includegraphics[scale=0.45]{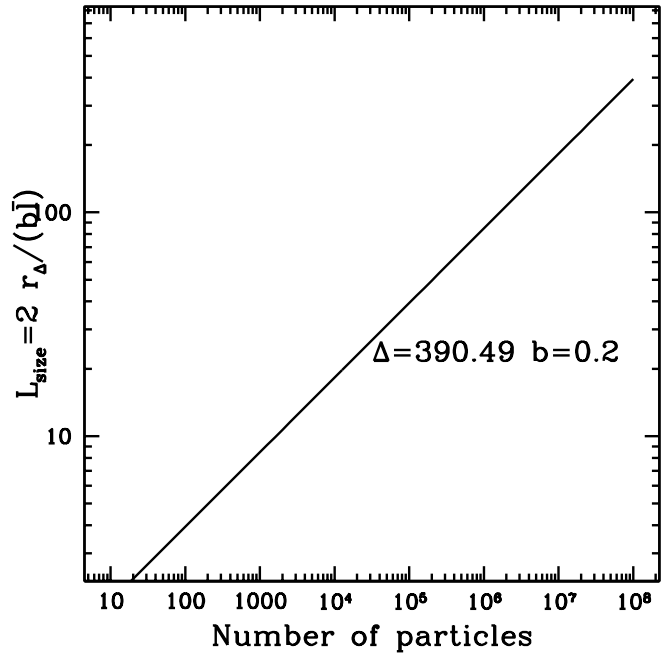}
\caption{ 
    The parameter $L_{\rm size}$ as a function of number of particles in a halo assuming halo concentration of $c_{180}=10$ and FOF linking length $b=0.2$. $L_{\rm size}$ defines the width of the FOF halo boundary. For halos with $L_{\rm size}\lesssim 10$ the FOF algorithm overestimates halo masses by $\gtrsim 10\%$ (see Figures~\ref{figwarren} and \ref{figfofmass}). 
  }
\label{fig10}
\end{figure}

We denote the fraction of spheres at any given radius that belong to
the infinite cluster by $f_{\rm accept}$. This fraction is simply the
ratio of the strength of the infinite cluster to the probability for
any point to belong to any sphere:
\begin{equation}
f_{\rm accept}=\frac{\pinf}{p}\,.
\label{eq:facc}
\end{equation}
In the right panel of Figure~\ref{fig8}, we show $\pinf$ as a function
of $p$ for the  NFW halo realizations. The line types and
colors are the same as in Figures~\ref{fig1} to \ref{fig4}. For $p\gg p_c$,
$f_{\rm accept}=1$ and $p\,f_{\rm accept}=p$. Near the percolation
threshold $p_c$, the fraction $f_{\rm accept}$ falls steadily from one
to zero in a way that depends upon the mean interparticle separation in the halo relative to the linking length.

We first investigate the strength of the infinite cluster, $\pinf$,
for $p>p_c$. In the left panel of Figure~\ref{fig9}, we show the
dependence of $\pinf$ on $p-p_c$ for $p>p_c$, obtained by analysing
the boundary of the NFW halo realizations identified by the FOF. The
bold grey line shows the percolation theory prediction given by
eq.~\ref{eq:pinf} with $\beta=0.43$. This prediction is in a very good
agreement with the results of the Monte Carlo simulations over an
order of magnitude in probability $p$ for the realizations with the
largest number of particles. In the right hand panel, we compare this
prediction to the results from the highest resolution halo.  We find
that percolation theory describes the behavior of the FOF boundary
for $p>p_c$ quite well. This explains why our empirical results for
the FOF boundary do not converge to a step function.

Note that the simple scaling of eq.~\ref{eq:pinf} predicts that
$\pinf\rightarrow 0$ as $p\rightarrow p_c$. This scaling, however, is
correct strictly for a uniform distribution of particles in an
infinite volume.  In contrast, realistic halos cover a finite volume
and have significant density gradients.  These effects change the
predictions of percolation theory \citep[e.g.][]{stauffer,rosso1986}.

For the standard case of percolation in an infinite volume with
uniform mean density, the connectivity length $\xi$ (expressed in
units of the sphere size or linking length) is the only scale in the
problem, and near the critical threshold $p_c$, the connectivity
length $\xi$ exhibits
critical scaling behavior, $\xi \propto |p_c - p|^{-\nu}$.  
In the more general case, other scales like the system size 
$L_{\rm size}$ or local scale length $s=p/|\nabla p|$ can be important
as well.  For example, in finite volumes percolation occurs when the
connectivity length becomes of order the system size, 
$\xi \approx {\cal O}(L_{\rm size})$, which occurs at a lower density
than infinite percolation.  The percolation threshold, therefore,
decreases as the system size decreases, and we can easily see that
setting $\xi \approx L_{\rm size}$ in Eqn.\ (\ref{xi_scale}) shows
that the finite-size threshold $\tilde{p}_c$ scales as 
\citep{stauffer}
\begin{equation}
\tilde{p}_c - p_c \propto L_{\rm size}^{-1/\nu}\,.
\end{equation}
Similarly, density gradients also modify the percolation transition.
Regions where the density is below the naive critical threshold,
$p<p_c$, can still be linked to regions above threshold, if the
connectivity length is of order the distance to the super-critical
region.  In other words, gradients will smear out the percolation
transition, by an amount that is straightforward to estimate.  If we
Taylor expand about the location where $p=p_c$, writing 
$p(x)=p_c + (\nabla p)x + \ldots$, then setting $x \approx \xi$
shows that the transition is smeared by a distance of roughly 
\begin{equation}
\xi \propto |p_c-p(\xi)|^{-\nu} = |p_c - p_c - (\nabla p) \xi|^{-\nu} 
 = |\nabla p\, \xi|^{-\nu} \Rightarrow \xi \propto |\nabla p|^{-\nu/(1+\nu)}.
\end{equation}
This corresponds to a width $\sigma_p$ in $p(x)$ such that
\begin{equation}
\sigma_p \propto |\nabla p|\, \xi \propto |\nabla p|^{1/(1+\nu)}\,.
\end{equation}
Thus, for non-uniform distributions, the density gradient results in a
much more gradual transition of $\pinf$ to zero, which extends to
$p<p_c$ \citep{rosso1986}, as illustrated in Fig.\ \ref{fig9}.

For realistic halos, both of the above effects (finite size and
density gradient) could be significant, but their importance must
diminish as the particle number in the halo increases. To judge the
importance of these effects for finite particle numbers,  
the quantity of interest is $L_{\rm size}=2 \, R_\Delta /
(b\bar{l})$,\footnote{The volume of the {\it system} enclosed by the boundary
$R_\Delta$ is equal to $4/3 \pi R_\Delta^3$ and the number of spheres of
radius $(b\bar{l})/2$ that can fit in this volume is equal to
$L_{\rm size}^3=8\,R_\Delta^3/(b \bar{l})^3$, which gives $L_{\rm size}=2 \,
R_\Delta / (b\bar{l})$.} where $R_\Delta$ is the threshold radius at
which the probability $p=p_c$.  In terms of the number of particles in
a FOF halo, $L_{\rm size}$ is given by
\begin{equation}
L_{\rm size}=\frac{2\,R_\Delta}{(b\bar{l})}=\frac{2}{b}\left(\frac{3
N_{\Delta}}{4 \pi \Delta}\right)^{1/3}.
\end{equation}
The analogous quantity for the gradient scale length will presumably
be of the same order as $L_{\rm size}$ for typical outer slopes in
halos, $|d\log\rho/d\log r| \sim 2-3$.  

In Figure~\ref{fig10}, we show $L_{\rm size}$ as a function of the
number of particles, $N_\Delta$, for halo realizations presented in 
\S~\ref{sec:od}. The FOF algorithm with a linking length parameter
$b=0.2$ selects an overdensity $\Delta=390.49$ for these halos with
concentration $c_{180}=10$. We note that even for $N_{\Delta}\approx
10000$, $L_{\rm size}\sim 10$. For such small values of 
$L_{\rm size}$, the threshold is significantly less than the infinite,
uniform density threshold, $\tilde{p}_c(L_{\rm size})<p_c$, meaning 
that the FOF algorithm joins particles
at radii corresponding to $p<p_c$ into the main halo.  This also leads to
an increase in the mass selected by FOF and a corresponding decrease
in the overdensity.

As we saw above, percolation theory predicts that the threshold value
for percolation scales with the size of the system (in units of the
linking length) as
$\tilde{p}_c - p_c \propto L_{\rm size}^{-1/\nu}$, \citep{stauffer}. 
This implies that the mass of halos selected
by the FOF algorithm will change as a function of $L_{\rm size}$ as 
\begin{equation}
\Delta M \propto \frac{\partial M}{\partial p} (\tilde{p}_c - p_c)
\propto  \left| \frac{\partial M}{\partial p}\right| L_{\rm size}^{-1/\nu}\,.
\label{eq:lsizescaling}
\end{equation}
To test this formula, we performed another set of
Monte-Carlo realizations of spherical halos. We assumed that the
particles follow a power law number density profile
\begin{equation}
n(r) \propto r^{-\alpha} \,.
\label{eq:nr}
\end{equation}
Following \citet{warren_etal06}, we arbitrarily normalized the halos to have
radius and mass equal to unity, $M=1$ and $R=1$, and used a linking length equal to
\begin{equation}
b\bar{l}=\left(\frac{N}{1.25}\right)^{-1/3}\,,
\end{equation}
where $N$ is the number of particles within $R=1$, to identify
halos. We generated halos with $\alpha \in
(1.5,1.75,2.0,2.25,2.5,2.75)$.  For each $\alpha$, we generated $10^3$
realizations each consisting of $100$, $500$ and $1250$ particles,
$100$ realizations each consisting of $10000$ and $80000$ particles,
ten realizations of $6.4\times 10^5$ particles, two realizations of
$6.4\times 10^6$ and one realization with $10^7$ particles. The value
of the radius $R_\Delta$ predicted using eq.~\ref{eq:ncrit} for these
halos is given by 
\begin{equation}
R_\Delta = \left[ \frac{4 \pi\, n_c}{1.25\,(3-\alpha)}
\right]^{-1/\alpha}\,.
\end{equation}
Note that $R_{\Delta}\ne R=1$ is the effective radius of the FOF boundary and we used the fact that $R=1$ in our model in the derivation of above equation. 
The corresponding value of $L_{\rm size}$ depends upon $\alpha$ and is
given by
\begin{equation}
L_{\rm size}=\frac{2 R_\Delta}{b\bar{l}} = 2\,\left(
\frac{N}{1.25}\right)^{1/3} \left( \frac{4 \pi\,
n_c}{1.25\,(3-\alpha)} \right)^{-1/\alpha}\,.
\end{equation}
Note that for increasing $\alpha$, the same number of particles, thus
correspond to a smaller value of $L_{\rm size}$. We would also like to
point out that the form of the density profile we chose in
Eq.\ref{eq:nr} above requires $\alpha<3$ to avoid the divergence in
mass at $r=0$. This does not imply that our formalism to correct the
masses of low resolution halos breaks down for $\alpha>=3$. As long
as $L_{\rm size}$, $\partial M/\partial p$ and $\alpha$ are calculated
appropriately at the boundary of the percolation threshold, our
formalism should work.

\begin{figure}[t]
\centering
\includegraphics[scale=0.9]{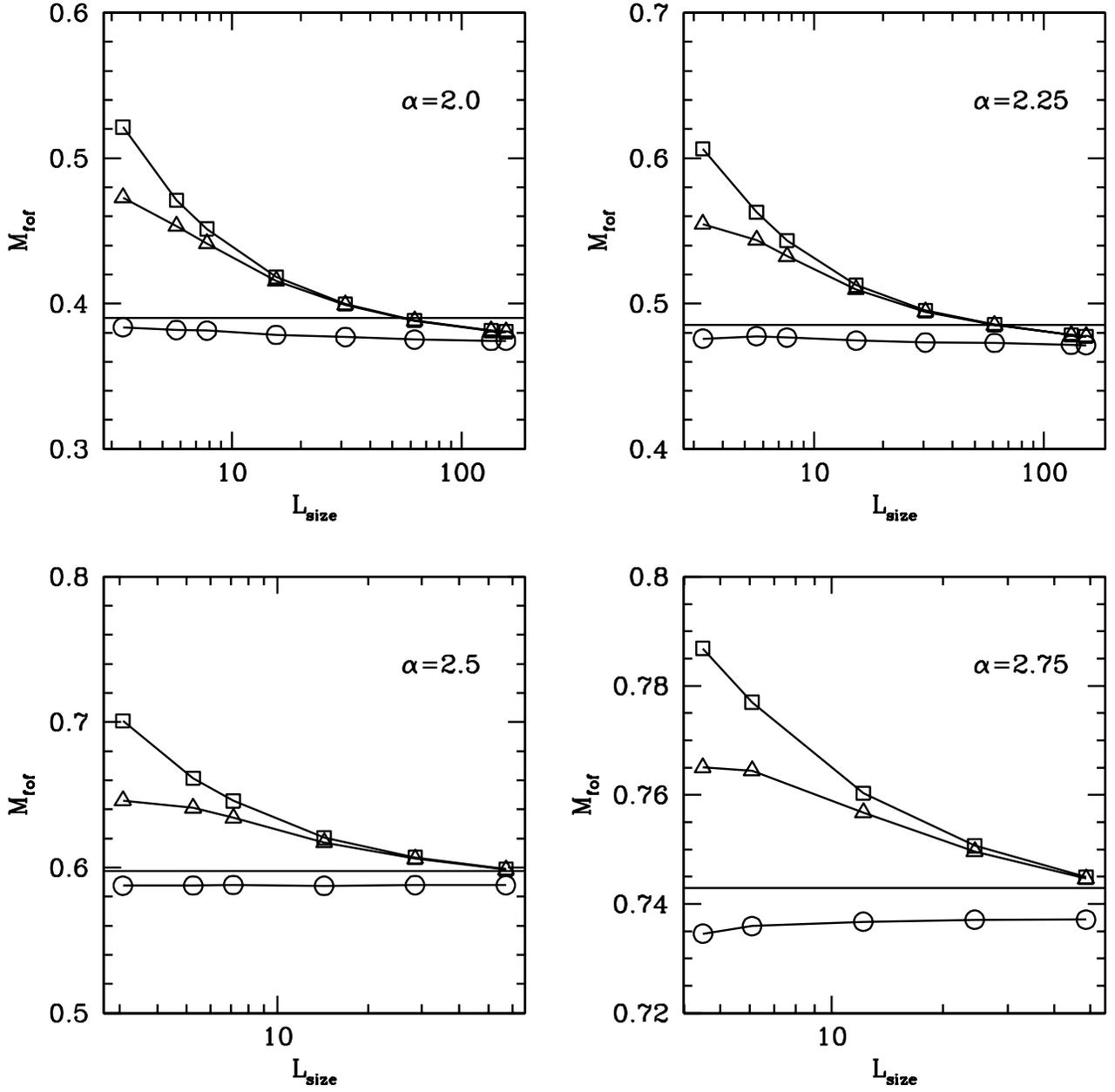}
\caption{The mass of the FOF halos characterized by different $L_{\rm size}$ for halos with power law density profiles $n(r)\propto r^{-\alpha}$. 
     Different panels correspond to different logarithmic slopes
    $\alpha$, as indicated in the legends.  Squares show the mass
    selected by the FOF algorithm ran on Monte Carlo realizations of
    halos, while triangles show masses corrected using empirical
    correction of \citet{warren_etal06}.  Open circles correspond to
    the FOF masses corrected using Eq.~\ref{eq:mcorrap}. The horizontal solid lines show the true mass  $M_{\Delta}$ for each halo model. }
\label{figfofmass}
\end{figure}

In each panel of Figure~\ref{figfofmass}, square symbols show the halo
mass of the main FOF halo as a function of $L_{\rm size}$ for
$\alpha=2.0,2.25,2.5$ and $2.75$. Other values of $\alpha$ give
similar results. The mass of the FOF halo asymptotes to its true value
as the number of particles with which the halo is sampled is
increased. This effect was first identified empirically 
by \citet{warren_etal06} and triangles show their proposed empirical 
correction. The figure shows, however, that this correction does not account
for the entire effect. The circles show the FOF masses corrected using
eq.~\ref{eq:lsizescaling} with a proportionality
constant of $0.22\,\alpha$ and $\nu=4/3$ \footnote{We have verified
with simple three dimensional gradient percolation experiments similar to
\citet{rosso1986} that $\nu=4/3$ in contrast to $\nu=0.88$ found for
three dimensions in case of uniform continuum percolation experiments.}:
\begin{equation}
M_{\rm fof}^{\infty} = M_{\rm fof}\, \left( 1 + 0.22\,\alpha
\,L_{\rm size}^{-1/\nu}\,\left|\frac{\partial\ln M_{\Delta}}{\partial
p}\right|
\right)^{-1}\,.
\label{eq:mcorrap}
\end{equation}
This correction almost entirely
eliminates the $L_{\rm size}$ dependence of the FOF-identified halo
mass. The circles thus represent the mass, $M^{\infty}_{\rm fof}$ that
would be selected by the FOF algorithm if it were run on a realization
with infinite number of particles. We note that for steeper density profiles (i.e., larger values of $\alpha$) a larger number of particles is required to converge to $M^{\infty}_{\rm
fof}$. 

As  was pointed out in
\S~\ref{sec:masses} and is clearly shown in Figure~\ref{figfofmass}, 
the mass $M_{\rm fof}^{\infty}$ is smaller than
the mass enclosed within an overdensity $\Delta$ given by
Eq.~\ref{eq:od} by a few percent. This is because the boundary
profile of the FOF halos is not a step function but has a specific shape
that can be approximately described by eq.\ref{eq:pinf} (see
Fig.~\ref{fig11}). This allows us to calculate an estimate
of the fraction $M_{\rm fof}^{\infty}/M_{\Delta}$ as 
\begin{equation}
\frac{M_{\rm fof}^{\infty}}{M_{\Delta}}= \frac{1}{\mu(c_\Delta)}\,\int^{c_\Delta}_{0}
f_{\rm accept}\, n(x) \,x^2 \drm x\,.
\end{equation}
Here the fraction $f_{\rm accept}$ and $P_\infty$ are given by eqs.~\ref{eq:facc} and ~\ref{eq:pinf}, respectively. As can be seen in
Figure~\ref{figwarren}, this boundary effect correction leads to values
of the masses that are very close to true mass $M_\Delta$.

In this appendix, we have presented a thorough analysis of the
boundary of the FOF halos in the context of percolation theory. We have
shown that percolation theory accurately predicts the shape of the
boundary of the FOF halos close to the density threshold for
percolation, at least for halos without significant amounts of
substructure (see \S~\ref{sec:masses}). We have also discussed how the finite number of particles with which a halo is sampled affects this boundary and have
found a percolation theory motivated formula to correct for this
dependence. Finally, we have also shown how the fraction
of mass identified by FOF in an infinite resolution halo relates to
the mass within a spherical overdensity given by eq.~\ref{eq:od}. 
These results provide a basis and theoretical interpretation for the empirical results presented in the main text of the paper. 

\end{document}